\documentstyle[12pt, epsf]{article}
\input math_macros.tex

\def\ref#1{$^{#1)}$}
\def\ha{\widehat{C}}
\def\wh{\widehat}
\def\wt{\widetilde{C}}
\def\wwt{\widetilde{Q}}
\def\sl{\slash}
\begin{document}
\begin{titlepage}
\begin{center}
\today     \hfill    LBL-35971 \\

\vskip .15in

{\large \bf Quantum Electrodynamics at Large Distances I: Extracting
the Correspondence-Principle Part.}
\footnote{This work was supported by the Director, Office of Energy
Research, Office of High Energy and Nuclear Physics, Division of High
Energy Physics of the U.S. Department of Energy under Contract
DE-AC03-76SF00098, and by the Japanese Ministry of Education,
 Science and Culture under a
Grant-in-Aid for Scientific Research (International Scientific Research Program
03044078).}

\vskip .1in
Takahiro Kawai\\

{\em Research Institute for Mathematical Sciences\\
Kyoto University\\
Kyoto 606-01 JAPAN\\}
\vskip .1in
Henry P. Stapp \\

{\em Lawrence Berkeley Laboratory\\
     University of California\\
     Berkeley, California 94720}

\end{center}

\vskip .05in

\begin{abstract}
The correspondence principle is important in quantum theory on both the
fundamental and practical levels: it is needed to connect theory to experiment,
and for calculations in the technologically important domain lying between the
atomic and classical regimes. Moreover, a correspondence-principle part of the
S-matrix is normally separated out in quantum electrodynamics in order to
obtain a remainder that can be treated perturbatively. But this separation, as
usually performed, causes an apparent breakdown of the correspondence
principle and the associated pole-factorization property. This breakdown is
spurious. It is shown in this article, and a companion, in the context of a
special case, how to extract a
distinguished part of the S-matrix that meets the correspondence-principle and
pole-factorization requirements. In a second companion paper the terms of the
remainder are shown to vanish in the appropriate macroscopic limits. Thus this
work validates the correspondence principle and pole factorization in quantum
electrodynamics, in the special case treated here,
and creates a needed computational technique.
\end{abstract}
\end{titlepage}
\renewcommand{\thepage}{\roman{page}}
\setcounter{page}{2}
\mbox{ }

\vskip 1in

\begin{center}
{\bf Disclaimer}
\end{center}

\vskip .2in

\begin{scriptsize}
\begin{quotation}
This document was prepared as an account for work sponsored by the United
States Government.  Neither the United States Government nor any agency
thereof, nor The Regents of the University of California, nor any of their
employees, makes any warranty, express or implied, or assumes any legal
liability or responsibility for the accuracy, completeness, or usefulness
of any information, apparatus, product, or process disclosed, or represents
that its use would not infringe privately owned rights.  Reference herein
to any specific commercial products process, or service by its trade name,
trademark, manufacturer, or otherwise, does not necessarily constitute or
imply its endorsement, recommendation, or favoring by the United States
Government or any agency thereof, or The Regents of the University of
California.  The views and opinions of authors expressed herein do not
necessarily state or reflect those of the United States Government or any
agency thereof of The Regents of the University of California and shall
not be used for advertising or product endorsement purposes.
\end{quotation}
\end{scriptsize}

\vskip 2in

\begin{center}
\begin{small}
{\it Lawrence Berkeley Laboratory is an equal opportunity employer.}
\end{small}
\end{center}

\newpage
\renewcommand{\thepage}{\arabic{page}}
\setcounter{page}{1}

\noindent{\bf 1. Introduction}

\vskip 9pt

The correspondence principle asserts that the predictions of quantum theory
become the same as the predictions of classical mechanics in certain
macroscopic limits. This principle is needed to explain why classical mechanics
works in the macroscopic domain. It also provides the logical basis
for using the language and concepts of classical physics to describe the
experimental arrangements used to study quantum-mechanical effects.

It is primarily within quantum electrodynamics that the correspondence
principle must be verified.  For it is quantum electrodynamics that controls
the properties of the measuring devices used in these experimental
studies.

In quantum electrodynamics the correspondence-principle has two
aspects. The first pertains to the electromagnetic fields generated by the
macroscopic motions of particles: these fields should
correspond to the fields generated under similar conditions
within the framework of classical electrodynamics. The second aspect
pertains to motion of the charged particles: on the macroscopic scale
these motions should be similar to the motions of charged
particles in classical electromagnetic theory.

The pole-factorization property is the analog in quantum theory of the
classical concept of the stable physical particle.
This property has been confirmed in a variety of rigorous
contexts$^{1,2,3}$ for theories in which the vacuum is the only state of zero
mass. But calculations$^{4,5,6}$ have indicated that the property fails
in quantum electrodynamics, due to complications associated with infrared
divergences. Specifically, the singularity associated with the propagation of a
physical electron has been computed to be not a pole. Yet if the mass of the
physical electron were $m$ and the dominant singularity of a scattering
function at $p^2=m^2$ were not a pole then physical electrons would, according
to theory, not propagate over laboratory distances like stable particles,
contrary to the empirical evidence.

This apparent difficulty with quantum
electrodynamics has been extensively studied$^{7,8,9}$, but not fully
clarified. It is shown here, at least in the context of a special case that is
treated in detail, that the apparent failure in quantum electrodynamics of
the classical-type spacetime behaviour of electrons and positrons in the
macroscopic regime is due to
approximations introduced to cope with infrared divergences.  Those divergences
are treated by factoring out a correspondence-principle
part, before treating the
remaining part perturbatively. It will be shown here, at least within the
context of the case examined in detail, that if an accurate
correspondence-principle part of the photonic field is factored
out then the required correspondence-principle and pole-factorization
properties do hold.  The apparent
failure of these latter two properties in the cited references are artifacts of
approximations that are not justified in the context of the calculation
of macroscopic spacetime properties: some factors $\exp ikx$ are replaced by
substitutes that introduce large errors for small $k$ but very large $x$.

The pole-factorization theorem, restricted to the
simplest massive-particle case, asserts the following:
Suppose the momentum-space scattering function for a process
$(1 + 2 \leftarrow 3 + x)$ has a nonzero connected component
$$
S_c(p_1, p_2; p_3, p_x) \delta^4(p_1 + p_2 - p_3 - p_x),
$$
and that the scattering function for a process $(x+4 \leftarrow 5+6)$ has a
nonzero connected component $$
S_c(p_x, p_4; p_5, p_6) \times \delta^4(p_x + p_4 - p_5 - p_6).
$$
Then, according to the theorem, the three-to-three scattering function
$$
S_c(p_1,p_2,p_4; p_3, p_4, p_5) \times \delta^4(p_1 + p_2 + p_4 - p_3 - p_5 -
p_6)
$$
must have the form
$$
{N(p_1, p_2, p_4; p_3, p_4, p_5)\over (p_1 + p_2 - p_3)^2 - m^2 +i0},
$$
where $m$ is the mass of particle $x$, and the residue of the pole is a (known)
constant times
$$
S_c(p_1, p_2; p_3, p_x) S_c(p_x, p_4; p_5, p_6)\times \delta(p_1 + p_2 + p_4 -
p_3 - p_5 - p_6),
$$
where $p_x = p_1 + p_2 - p_3 = p_5 + p_6 - p_4$.

The physical significance of this result arises as follows.
Suppose we form wave packets $\psi_i(p_i) (i = 1, ..., 6)$ for the six
(external)
particles of the three-to-three process.
Let these momentum-space wave packets be nonzero at a set of values
$\overline{p}_i$
such that
$$
\overline{p}_1 + \overline{p}_2 - \overline{p}_3 = \overline{p}_x =
\overline{p}_5 + \overline{p}_6 - \overline{p}_4
$$ where $(\overline{p}_x)^2 = m^2$.
Suppose the corresponding free-particle coordinate-space wave packets
$\tilde{\psi}_i(x_i, t)$ for these six particles are all large at the origin of
spacetime.
Now translate the wave packets of particles 1,2, and 3 by the spacetime
distance
$\lambda \overline{p}$, and let $\lambda$ tend to infinity.
Then $\lambda ^{3/2}$ times this $3 \to 3$ transition amplitude must, according
to
the theorem, tend to a limit that is a (known) constant times the product of
the
two scattering amplitudes,
$$
A_D = \int d^3p_1 \int d^3 p_2 \int d^3 p_3 \psi^*_1(p_1) \psi^*_2(p_2)
\psi_3(p_3)
S_c(p_1, p_2; p_3, \overline{p}_x)
$$
and
$$
A_P = \int d^3p_4 d^3p_5 d^3 p_6 \psi^*_4(p_4) \psi_5(p_5) \psi_6(p_6)
S_c(\overline{p}_x, p_4; p_5, p_6).
$$

This result has the following physical interpretation: the transition
amplitude $A_P$ is the amplitude for {\it producing} a particle $x$ of
momentum $\overline{p}_x$, and the amplitude $A_D$ is the amplitude for
{\it detecting} this particle.
The fall-off factor $\lambda^{- 3/2}$ becomes $\lambda^{-3}$ when one passes
from
amplitudes to probabilities, and this factor $\lambda^{-3}$ is what would be
expected on purely geometric grounds in classical physics,
if the intermediate particle $x$ produced by the production process, and
detected by the detection process travelled, in the asymptotic regime,
on a straight line in spacetime with four-velocity
$v_x = p_x/m \simeq \overline{v}_x = \overline{p}_x/m$.

This $\lambda^{-3}$ fall-off property is also what is observed empirically for
both neutral and charged particles travelling over large distances in free
space. On the other hand, computations$^{4,5,6}$ in QED
have shown that if quasi-classical parts are factored off in the usual
momentum-space manner then in the remainder the singularies associated with
the propagation of charged particles have, instead of the pole form
$(p^2 -m^2 +i0)^{-1}$, rather a form $(p^2 - m^2 +i0)^{-1-\beta}$, where
$\beta$ is nonzero and of
order $\alpha \approx 1/137$. Such a form would entail that electrons and
positrons would not behave like stable particles: they would evoke weaker and
weaker detection signals (or seems to disappear) for $\beta < 0$, or evoke
stronger and stronger detection signals for $\beta > 0$, as their distance
from the source increases.

Such effects are not observed empirically.
Hence $\beta$ must be zero (or at least close to zero), in apparent
contradiction to
the results of the cited QED calculations.

For the idealized case in which all particles have nonzero mass the
pole-factorization theorem has been proved in many ways.
The simplest ``proof'' is simply to add up all of the Feynman-graph
contributions
that have the relevant pole propagator $(p^2 - m^2 + i0)^{-1}$, and observe
that
the residue has the required form.
Proofs not relying on perturbation theory have been given in the frameworks of
quantum field theory$^1$,
constructive field theory$^2$, and  $S$-matrix theory$^3$.

In quantum electrodynamics if the particle $x$ is {\it charged} then at least
one other charged particle must either enter or leave each of the two
subprocess, in order for charge to be conserved.
If there is a deflection of this charged particle in either of these two
subprocesses then bremsstrahlung radiation will be emitted by that process.
As $\lambda \to \infty$ the number of photons radiated will tend to infinity.
Thus in place of the two simple $2 \to 2$ sub-processes considered in the
example discussed above one must include in QED the bremsstrahlung photons
radiated at each of the two subprocesses.

Bremsstruhlung photons were in fact taken into account in the earlier cited
works$^{1,2,3}$.
However,  in those works it was assumed, in effect,  that all of these
photons were emitted from a neighborhood of the origin in spacetime.
This imprecision in the positioning of the sources of the bremsstrahlung
radiation arose from the use of a basically momentum-space approach.

It is clear that coordinate-space should provide a more suitable
framework for accurately positioning the sources of the radiated photons.
Indeed, it turns out that it is sufficient to place the sources of the
(real and virtual) bremsstrahlung photons at the physically correct positions
in coordinate space in order to establish the validity of a
pole-factorization property in QED, at least in the special case that we study
in detail in this paper.

Examination of the work of Kibble$^4$ shows that there is, in the case he
treated, also another problem. In that case some of the charged-particle lines
extend to plus or minus infinity. At one point in the calculation, a factor
$(p_\mu /p\cdot k)(e^{ipx_1} -e^{ikx_2})$ initially associated with such a
line,
where $x_1$ and $x_2$ represent the two ends of the charged particle line, is
replaced by a single one of the two terms: the other term, corresponding to the
point $x_i =\infty$, is simply dropped.
Yet dropping this term alters the character of the behavior at $k=0$: the
original product of this form with $k^\mu$ tends to zero as $k$ vanishes,
but to plus or minus unity if a term is dropped.

It turns out that this treatment of the contributions corresponding to points
at
infinity leads to serious ambiguities.$^{7}$
To avoid such problems, and keep everything finite and well defined in the
neighborhood of $k =0$, we shall consider the case of a ``confined charge'';
i.e., a case in which a charge travels around a closed loop in spacetime,
in the Feynman sense: a backward moving electron is interpreted as a forward
moving positron.
In particular, we shall consider an initial graph in which the charge
travels around a closed triangular loop $L(x_1, x_2, x_3)$ that has vertices at
spacetime points $x_1$, $x_2,$ and $x_3$. These three vertices represent points
where ``hard'' photons interact. (Actually, each $x_i$ will correspond to a
{\it pair} of hard-photon vertices, but we shall, in this introduction, ignore
this slight complication, and imagine the two hard photons to be attached to
the same
vertex of the triangle.) We must then consider the effects of inserting
arbitrary numbers of ``soft photon'' vertices into this hard-photon triangle
in all possible ways.
The three hard-photon vertices $x_i$ are held fixed during most of the
calculation. At the end one must, of course, multiply this three-point
coordinate-space scattering function by the coordinate-space
wavefunctions of the external particles connected at these three spacetime
points, and then integrate over all possible values of $x_1, x_2$, and $x_3$.

As in the two-vertex example given above, we are interested  in the
behavior in the limit in which $(x_1, x_2, x_3)$ is replaced by
$(\lambda x_1, \lambda x_2, \lambda x_3)$ and $\lambda$ tends to infinity.
The physically expected fall-off rate is now $(\lambda^{-3/2})^3$, with one
geometric fall-off factor
$\lambda^{- 3/2}$ for each of the three intermediate charged-particle lines.

This $\lambda^{-9/2}$ fall off is exactly the coordinate-space fall-off that
arises from a Feynman function corresponding to graph consisting of external
lines connected to the three vertices of a triangle of internal lines.
The singularity in momentum space corresponding to such a simple triangle
graph is log $\varphi$, where
$$
\varphi = \varphi (q_1, q_2, q_3)=0
$$
is the so-called Landau-Nakanishi (or, for short, Landau) triangle-diagram
singularity
surface. Here the $q_i$ are the momenta entering the three vertices, and they
are subject to the momentum-energy conservation law $q_1+q_2+q_3=0$.

In close analogy to the single-pole case discussed earlier, the discontinuity
of the full scattering function across this log $\varphi$ surface at
$\varphi =0$ is, in theories with no massless particles, a (known) constant
times a product of three scattering functions, one corresponding to each of
the three vertices of the triangle:
$$
 \mbox{disc} S|_{\varphi =0} = \mbox{const.} \times S_1S_2S_3.
$$

It will be shown in these papers  that this formula for the discontinuity
around the triangle-diagram singularity surface $\varphi = 0$ holds also
in quantum
electrodynamics to every order of the perturbative expansion in the
nonclassical
part of the photon field. The situation is more complicated than in the
massive-particle case because now an infinite number of singularities
of different types all coincide with $\varphi =0$. It will be shown
that many  of these do not contribute to the discontinuity at  $\varphi = 0$,
because the associated discontinuities contain at least one  full power
of $\varphi$, and that all of the remaining contributions are parts of the
discontinuity function given above.

Another complication is that an infinite number of photons are radiated from
each of the three vertices of the triangle. In our treatment these photons are
contained in the (well-defined) classical part of the photon field. The
contributions from these photons depend on the locations of the vertices $x_i$,
and are incorporated {\it after} the transformation to
coordinate space.

This focus on the triangle-graph process means that we are dealing here
specifically with the charge-zero sector. But the scattering functions for
charged sectors can be recovered by exploiting the proved pole-factorization
property. It is worth emphasizing, in this connection, that a straight-forward
application of perturbation theory in the triangle-graph case does {\it not}
yield the pole-factorization property, even though the triangle graph
represents a process in the charge-zero sector. Just as in the charged sectors,
it is still necessary to separate out the part corresponding to the classical
photons. If one does not, then the first-order perturbative term gives a
singularity of the form$^{10}$  $(log\varphi)^2$, instead of the physically
required
form $log\varphi$. It is worth emphasizing that we do not neglect ``small''
terms in denominators, but keep everything exact. Indeed, it is important that
we do so, because these small terms are essential to the validity of law of
conservation of charge, which we use extensively.

In the foregoing discussion we have focussed on the pole-factorization-theorem
aspects of our work. But the paper contains much more.
It provides the mathematical machinery needed to apply quantum electrodynamics
in the mesoscopic and macroscopic regimes where charged particles move between
interaction regions that are separated by distances large enough for the
long-distance particle-type behaviours of these  particles to begin to
manifest themselves.
That is, this paper establishes a formalism that allows quantum
electrodynamics  to be accurately applied to the transitional domain lying
between the quantum and classical regimes.The machinery displays in a
particularly simply and computationally useful form the infrared-dominant
``classical'' part of the electromagnetic field, while maintaining good
mathematical control over the remaining ``quantum'' part.

This work is based on the separation defined in reference 11  of the
electromagnetic interaction operator into its ``classical'' and ``quantum''
parts. This separation is made in the following way. Suppose we first make a
conventional energy-momentum-space separation of the (real and virtual photons)
into ``hard'' and
``soft'' photons, with hard and soft photons connected at ``hard'' and ``soft''
vertices, respectively. The soft photons can have small energies and momenta
on the scale of the electron mass, but we shall not drop any ``small'' terms.
Suppose a charged-particle line runs from a hard
vertex $x^-$ to a hard vertex $x^+$. Let soft photon $j$ be coupled into this
line at point $x_j$, and let the coordinate
variable $x_j$ be converted by Fourier transformation to the associated
momentum variable $k_j$.
Then the interaction operator $-ie\gamma_{\mu_j}$ is separated into its
``classical'' and ``quantum'' parts by means of the formula
$$
-ie \gamma_{\mu_j}= C_{\mu_j} + Q_{\mu_j}, \eqno(1.1)
$$
where
$$
C_{\mu_j} = -ie{z_{\mu_j}\over z\cdot k_{j}} \slash{k}_j, \eqno(1.2)
$$
and $z=x^+ - x^-$.

This separation of the interaction allows a corresponding separation of soft
photons into ``classical'' and ``quantum'' photons: a ``quantum'' photon has a
quantum coupling on at least one end; all other photons are called
``classical''
photons. The full contribution from all classical photons is represented in an
extremely neat and useful way. Specialized to our case of a single
charged-particle loop $L(x_1, x_2, x_3)$ the key formula reads
$$
F_{op}(L(x_1,x_2, x_3))=:U(L(x_1,x_2,x_3)) F'_{op} (L(x_1,x_2,x_3)):.\eqno(1.3)
$$
Here $F_{op} (L(x_1, x_2, x_3))$ is the Feynman {\it operator} corresponding
to the sum of contributions from {\it all}
photons coupled into the charged-particle loop $L(x_1, x_2, x_3)$, and
$F_{ op}'(L(x_1, x_2, x_3))$ is the analogous operator if
all contributions from classical photons are excluded.
The operators $F_{ op}$ and $F'_{ op}$ are both normal ordered operators: i.e.,
they are operators in the asymptotic-photon Hilbert space, and the destruction
operators of the incoming photons stand to the right of the creation
operators of outgoing photons. On the right-hand side of $(1.3)$ all of the
contributions corresponding to classical photons are included in
the unitary-operator factor $U(L)$ defined as follows:
$$
U(L) = e^{<a^*\cdot J(L)>} e^{-\half <J^*(L)\cdot J(L)>}
      e^{-<J^*(L)\cdot a>}e^{i\Phi (L)}. \eqno(1.4)
$$
Here, for any $a$ and $b$, the symbol $<a\cdot b >$ is an abbreviation for the
integral
$$
<a\cdot b>\equiv \int {d^4k\over (2\pi)^4} 2\pi \theta (k_0)\delta (k^2)
a_\mu(k)(-g^{\mu\nu} )b_\nu(k),\eqno(1.5)
$$
\noindent and $J(L,k)$ is formed by integrating $\exp ikx$ around the loop $L$:
$$
J_\mu(L,k) \equiv \int_L dx_\mu e^{ikx}.\eqno(1.6)
$$
This classical current $J_{\mu}(L)$ is conserved:
$$
k^\mu J_\mu (L, k) =0. \eqno(1.7)
$$
The $a^*$ and $a$ in $(1.4)$ are photon creation and destruction operators,
respectively, and
$\Phi (L)$ is the classical action associated with the motion of a charged
classical particle along the loop $L$:
$$
\Phi (L) = {(-ie)^2\over 8\pi} \int_L dx'_{\mu} g^{\mu\nu} \int_L dx''_\nu
\delta((x' - x'')^2)\eqno(1.8)
$$
The operator $ U(L)$ is {\it pseudo} unitary if it is written in explicitly
covariant form, but it can be reduced to a strictly
unitary operator using by $(1.7)$ to eliminate all but the two
transverse components of $a_\mu (k),a^*_\mu (k), J_\mu(k)$, and $J^*_\mu (k)$.

The colons in (1.3) indicate that the creation-operator parts of the normal-
ordered operator $F'_{op}$ are to be placed on the left of $U(L)$.

The unitary operator $U(L)$ has the following property:
$$
U(L)|vac > = |C(L) >. \eqno(1.9)
$$
Here $|vac>$ is the photon vacuum, and $|C(L)>$ represents the normalized
coherent state corresponding to
the classical
electromagnetic field radiated by a charged classical point particle
moving along the closed spacetime loop $L$,
in the Feynman sense.

The simplicity of (1.3) is worth emphasizing: it says that the complete effect
of all classical photons is contained in a simple multiplicative factor that is
independent of the quantum-photon contributions: this factor is a well-defined
unitary operator that depends only on the (three)
hard vertices $x_1, x_2$, and $x_3$. It is independent of the remaining
details of $F'_{op}(L(x_1, x_2, c_3))$, even though the
classical couplings are originally interspersed in all possibly ways among the
quantum couplings that appear in $F'_{op}(L(x_1, x_2, x_3))$.
The operator $U(L)$ supplies the classical bremsstrahlung-radiation photons
associated with the deflections of the charged particles that occur at the
three vertices, $x_1,x_2,$ and $x_3$.

Block and Nordsieck$^{12}$ have already emphasized that the infrared
divergences arise from the classical aspects of the elecromagnetic field.
This classical component is exactly supplied by the factor $U(L)$.
One may therefore expect the remainder $F'_{op} (L(x_1, x_2, x_3))$ to be
free of infrared problems: if we
transform $F'_{op}(L(x_1,x_2,x_3))$ into momentum space, then it should satisfy
the usual pole-factorization property. A primary goal of this work is to show
that this pole-factorization property indeed holds. To recover the physics one
transforms $F'_{op}$ to coordinate space,
and then incorporates the real and virtual classical photons by using
$1.3$ and $1.4$.

The plan of the paper is as follows.
In the following section  2 rules are established for writing down the
functions of interest directly in momentum space.
These rules are expressed in terms of operators that act on
momentum--space Feynman functions and yield momentum--space functions, with
classical or quantum interactions inserted into the charged-particle lines
in any specified desired order.

It is advantageous always to sum together the contributions corresponding
to all ways in which a photon can couple with C--type coupling into each
individual side of the triangle graph $G$. This sum can be expressed as a
sum of just two terms. In one term the photon is coupled at one endpoint,
$x^+$, of this side of $G$, and in the other term the photon is coupled into
the other end point, $x^-$, of this side of $G$.
Thus all C--type couplings become converted into couplings at the hard--photon
vertices of the original graph $G$.

This conversion introduces an important
property. The charge--conservation (or gauge) condition $k^\mu J_\mu =0$
normally does not hold in quantum electrodynamics for individual graphs:
one must sum over all ways in which the photon can be inserted into the graph.
But in the form we use, with each quantum vertex $Q$ coupled into the interior
of a line of $G$, but each classical vertex $C$ placed at a hard--photon
vertex of $G$, the charge--conservation equation (gauge invariance) holds for
each vertex separately: $k^\mu J_\mu =0$ for each vertex.

In section 3 the modification of the charged--particle propagator caused by
inserting a single quantum vertex $Q_\mu$ into a charged-particle line is
studied in detail. The resulting (double) propagator is re--expressed as
a sum of three terms.
The first two are ``meromorphic'' terms having poles at $p^2=m^2$ and $p^2
= m^2-2pk -k^2$, respectively, in the variable $p^2$.
Because of the special form of the quantum coupling $Q_\mu$ each residue is of
first order in $k$, relative to what would have been obtained with the usual
coupling $\gamma_\mu$. This extra power of $k$ will lead to the infrared
convergence of the residues of the pole singularities.

The third term is a nonmeromorphic contribution.
It is a difference of two logarithms. This {\it difference} has a power of
$k$ that renders the contribution infrared finite.

In section 4 the results just described are used to study the function
corresponding to a graph $g$ that is formed by inserting into the triangle
graph $G$ a single quantum photon that has Q--type interactions at each
end.
In order to treat in a rigorous way the contribution from the neighborhood
of the point $k=0$ we introduce polar coordinates $k=r\Omega, \ \Omega
\tilde{\Omega} \equiv \Omega^2_0 + \vec{\Omega}\cdot \vec{\Omega}=1$.
For the meromorphic contributions it is found that  the integrand
of the integral that defines the residue behaves like $rdr$ near the end point
$r=0$, and that the compact domain of  integration
in the variable $\Omega$ can be distorted away from all singularities.
This shows that there is no infrared divergence. The two meromorphic
contributions from each end of the photon line lead to four contributions to
$F(D')$.
One of them gives the normal log $\varphi$ singularity on  the Landau
triangle--diagram
surface $\varphi =0$, and the other three give weaker singularities.
The contributions from the nonmeromorphic contributions also give weaker
singularities.

The aim of the remaining sections is basically to prove that the analogous
results hold for all graphs $g$ constructed from the original triangle
graph $G$ by the addition of any number of quantum--photon lines. In the
process of proving this, we construct the foundation of an efficient general
machinery for computing, in quantum electrodynamics, the physical-region
singularity structure, or, equivalently, an accurate representation of the
large-distance spacetime behavior.

In section 5 we examine the {\it generalized propagator} that
 corresponds to charged-particle propagation
between two hard--photon vertices $x$ and $y$ with an arbitrary number of
Q--type insertions.
The meromorphic part is exhibited explicitly:
there is one pole term for each of the original energy denominators.
The residues factorize, and each of the two factors (unless it is unity) has
{\it one} factor of $k_i$ beyond what would occur if the couplings were the
original $\gamma_\mu$ couplings.
This single extra factor of $k_i$ in each residue factor  will lead
to infrared convergence of the meromorphic parts.

This infrared convergence result, for any graph $g$ obtained by inserting
a  set of internal quantum photons into the triangle graph $G$,
is proved in sections 6 and 7, subject to the assumption that, in analogy to
what occurred in the simple case treated in section $4$, the $\Omega$
contours can be distorted so as to avoid all singularities of the residue
factors. This distortion assumption reduces the problem to that of counting
powers
of $r$. However, it is not sufficient merely to count overall powers of $r$.
One must show that, for every possible way in which the variables $k_i$
can tend to zero, there is convergence of every sub-integral.
Our proof that this convergence property holds can be regarded as a
systematization and confirmation of the
argument for infrared convergence given by Grammer and Yennie$^{13}$.
The problem is non-trivial because for every $n>0$ there are terms with
$n$ factors of the form $d^4 k_i / D_i$, where the denominator $D_i$ is of
fourth order in the $k_i$, for small $k_i$, but there are at most six
extra numerator factors $k_i$: the problem is to show that these few factors
$k_i$ are enough to block all of the $4n$ potential logarithmic divergences.
The problem is basically a topological one.

Section 8 gives a desciption of a result that asserts that the contours
in $\Omega_i$-space can be distorted away from all singularities of the
residue factors and photon propagators . The proof of this result is given
in a companion paper$^{14}$.

In section 9 the results of the earlier sections are gathered together and
extended to give the result that singularities on the triangle-diagram surface
coming from the meromorphic parts of the contributions arising from the
quantum photons are no stronger than $log\varphi$. References are made to a
second companion paper$^{15}$, which proves pertinent properties of some
integrals that occur in this work. In sections 10 and 11 the similar results
for the nonmeromorphic parts are obtained. Section 12 gives a comparison of
the present work to recent related works.

To get papers of manageable size we have separated the work into three
articles, of which this is the first. The second$^{14}$ contains the proof
that in the pole--decomposition functions that we have separated out
the contours in the angular variables $\Omega_i$ can be distorted away from
all singularities, with the exception of three Feynman
denominators, one from each side $s$ of the original triangle graph $G$, and
the
end points of the radial integrations. This means that each of the
distinguished contributions corresponding to a {\it separable} graph $g$ is
essentially the same as the usual triangle-graph function, multiplied by a
bounded function of the variables $r_i$ and $\Omega_i$ and integrated over a
compact domain in these variables. The factor $U(L)$ supplies the
quantum analog of the appropriate classical electromagnetic field.

The final but crucial point is then to show that the remaining parts, which
are specified by compactly expressed integrals, give contributions that
tend to zero in the macroscopic regime, relative to distinguished part
discussed above, which meets the correspondence--principle and
pole--factorization requirements. The required properties of these integrals
are established in the third paper$^{15}$. Our original report$^{16}$ contains
all three parts in one place.
\newpage
\noindent
{\bf 2.  Basic Momentum--Space Formulas}

The separation of the soft--photon interaction into its quantum and
classical parts is defined in Eq. (1.1).
This separation is defined in a mixed representation in which hard
photons are represented in coordinate space and soft photons are
represented in momentum space.
In this representation one can consider a ``generalized propagator''.
It propagates a charged particle from a hard--photon vertex $y$ to a
hard--photon vertex $x$ with, however, the insertion of soft--photon
interactions.

Suppose, for example, one inserts the interactions with two soft photons
of momenta $k_1$ and $k_2$ and vector indices $\mu_1$ and $\mu_2$.
Then the generalized propagator is
$$
\eqalignno{
P_{\mu_1, \mu_2} &(x,y; k_1, k_2)\cr
&= \int {d^4p\over (2\pi )^4} e^{-ipx + i(p+k_1+k_2)y}\cr
&\times {i\over \slash{p}-m+i0}\gamma_{\mu_1}{i\over
\slash{p}+\slash{k}_1-m+i0}\gamma_{\mu_2}{i\over
\slash{p}+\slash{k}_1+\slash{k}_2-m+i0}.&(2.1)\cr}
$$
The generalization of this formula to the case of an arbitrary number of
inserted soft photons is straightforward.
The soft--photon interaction $\gamma_{\mu_j}$ is separated into its parts
$Q_{\mu_j}$ and $C_{\mu_j}$ by means of (1.1), with the $x$ and $y$ defined
as in (1.3).

This separation of the soft--photon interaction into its quantum and
classical parts can be expressed also directly in momentum space.
Using (1.2) and (1.3), and the familiar identities
$$
{1\over \slash{p}-m} \slash{k}
{1\over \slash{p} + \slash{k} - m} = {1\over
\slash{p}-m} - {1\over \slash{p}+ \slash{k}-m},\eqno(2.2)
$$
and
$$
\left( - {\partial\over \partial p^\mu}\right)
{1\over \slash{p}-m} = {1\over
\slash{p}-m} \gamma_\mu {1\over \slash{p}-m},\eqno(2.3)
$$
one obtains for the (generalized) propagation from $y$ to $x$, with a
single classical interaction inserted, the expression (with the symbol
$m$ standing henceforth for $m-i0)$
\newpage
$$
\eqalignno{
P_\mu(x,y; C,k)
&= \int {d^4p\over (2\pi )^4}
\left(
{i\over \slash{p}-m}\slash{k}
{i\over \slash{p}+\slash{k}-m}\right)
{z_\mu\over zk+io} e^{-ipz+iky}\cr
&= \int {d^4p\over (2\pi )^4}
\left( {i\over \slash{p}-m}
\slash{k}
{i\over \slash{p}+\slash{k}-m}\right)
{1\over zk+io}
\left(
{i\partial\over \partial p^\mu}\right)
e^{-ipz+iky}\cr
&= \int {d^4p\over (2\pi )^4}
e^{-ipz+iky}
{1\over zk+io}
\left( -i{\partial\over \partial p^\mu}\right)
\left( {i\over \slash{p} -m}
\slash{k}{i\over
\slash{p}+\slash{k}-m}\right)\cr
&= \int{d^4p\over (2\pi )^4}
e^{-ipz+iky}
\mathop{\lim}_{\epsilon \to 0}(-i)
\int^\infty_0 d\lambda
e^{i\lambda (zk+i\epsilon )}\cr
&\hbox{\hskip .25in}\times \left(
-i{\partial\over\partial p^\mu} \right)
\left({i\over \slash{p}-m}
\slash{k}{i\over \slash{p}+\slash{k}-m}\right)\cr
&=\int {d^4p\over (2\pi )^4}
\mathop{\lim}_{\epsilon \to 0} (-i)
\int^\infty_0
d\lambda \ e^{-i(p-\lambda k)z+iky - \epsilon \lambda}\cr
&\hbox{\hskip .25in}
\times \left( -i{\partial\over \partial p^\mu}\right)
\left(
{i\over \slash{p}-m}
\slash{k}{i\over \slash{p}+\slash{k}-m}\right)&(2.4a)\cr
&= \int {d^4p\over (2\pi )^4}
\mathop{\lim}_{\epsilon \to 0} (-i)
\int^\infty_0 d\lambda \ e^{-i
(p - \lambda k)z+iky - \epsilon \lambda}\cr
&\hbox{\hskip .25in}\times {\partial\over \partial p^\mu}
\left( {i\over \slash{p}-m} -
{i\over \slash{p}+\slash{k}-m}\right)\cr
&= \int {d^4p\over (2\pi )^4}
\mathop{\lim}_{\epsilon \to 0}  (-i)
\int^\infty_0 d\lambda \ (e^{-i(p-\lambda k)z}-
e^{-i (p-k- \lambda k)z})e^{-\epsilon \lambda}\cr
&\hbox{\hskip .25in}\times e^{iky}{\partial\over\partial p^\mu}
\left( {i\over \slash{p}-m}\right)\cr
&= \int{d^4p\over (2\pi )^4} \mathop{\lim}_{\epsilon\to 0} (-i)\int^1_0
d\lambda \ e^{-i(p-\lambda k)z} e^{-\epsilon\lambda}
\times e^{iky}{\partial\over\partial
p^\mu}\left({i\over\slash{p}-m}\right)\cr
&= \int {d^4p \over (2\pi )^4} e^{-ipz+iky}
\int^1_0
d\lambda\left(-i {\partial\over\partial p^\mu}\right)
\left({i\over \slash{p}+\lambda\!\slash{k}-m}
\right)&(2.4b)\cr
&=\int {d^4p\over (2\pi )^4}  e^{-ipz+iky}
\int^1_0 d\lambda
\left({i\over \slash{p}+\lambda\!\slash{k}-m}
\gamma_\mu
{i\over \slash{p}+\lambda\!\slash{k}-m}\right).\cr
&&(2.4c)\cr}
$$

Comparison of the result (2.4b) to (2.1) shows that the result in
momentum space of inserting a single quantum vertex $j$ into a
propagator $i(\slash{p}-m)^{-1}$ is produced by the action of the
operator
$$
\widehat{C}_{\mu_j} (k_j)= \int^1_0 d\lambda_j O(p\to p+
\lambda_j k_j)\left(-i{\partial\over \partial p^{\mu_j}}\right)\eqno(2.5)
$$
upon the propagator $i(\slash{p}- m)^{-1}$ that was present
{\it before} the insertion of the vertex $j$.
One must, of course, also increase by $k_j$ the momentum entering the
vertex at $y$.
The operator $O(p\to p+\lambda_jk_j)$ replaces $p$ by $p+\lambda_jk_j$.

Suppose that there were already a soft--photon insertion on the charged
--particle line $L$ so that the propagator before the insertion of
vertex $j$ were
$$
P_{\mu_1}(p; k_1) ={i\over\slash{p}-m}\gamma_{\mu_1}{i\over \slash{p}
+\slash{k}_1-m}.\eqno(2.6)
$$
And suppose the vertex $j$ is to be inserted in all possible ways into
this line (i.e., on both sides of the already--present vertex 1).
Then the same argument as before, with (2.2) replaced by its
generalization$^{9}$
$$
\eqalignno{
{1\over \slash{p}-m}&\slash{k}_j
{1\over \slash{p} + \slash{k}_j-m}
\gamma_{\mu_1} {1\over \slash{p}+\slash{k}_j+\slash{k}_1-m}\cr
&+ {1\over \slash{p}-m}
\gamma_{\mu_1}{1\over \slash{p} + \slash{k}_1-m} \slash{k}_j
{1\over \slash{p}+\slash{k}_j +
\slash{k}_1-m}\cr
&= {1\over \slash{p}-m}
\gamma_{\mu_1}{1\over \slash{p}+\slash{k}_1 -m}\cr
&- {1\over \slash{p} + \slash{k}_j -m}
\gamma_{\mu_1}{1\over \slash{p}+\slash{k}_j + \slash{k}_1-m},&(2.7)\cr}
$$
shows that the effect in momentum space is again given by the operator
$\widehat{C}_{\mu_J}(k_j)$ defined in (2.5).

This result generalizes to an arbitrary number of inserted classical
photons, and also to an arbitrary generalized propagator: the
momentum--space result of inserting in all orders into any generalized
propagator $P_{\mu_1 \cdots \mu_n} (p; k_1, \cdots k_n)$ a set of $N$
classically interacting photons with $j= n+1, \cdots, n+N$ is
\newpage
$$
\eqalignno{
&\prod^{n+N}_{j=n+1}
\widehat{C}_{\mu_j}(k_j)
P_{\mu_1, \cdots , \mu_n}
(p; k_1, \cdots , k_n)
=\int^1_0 \ldots \int^1_0 d\lambda_{n+1}\ldots d\lambda_{n+N}
\prod^{N}_{j=1} \left( -i{\partial\over \partial p^{\mu_{n+j}}}\right)\cr
&\hbox{\hskip.25in} P_{\mu_1, \cdots, \mu_n} (p+a; k_1, \cdots , k_n)&(2.8)\cr}
$$
where $a= \lambda_{n+1} k_{n+1} + \cdots + \lambda_{n+N} k_{n+N}$.
The operations are commutative, and one can keep each $\lambda_j=0$
until the integration on $\lambda_j$ is performed.

To obtain the analogous result for the quantum interactions we introduce
the operator $\widehat{D}_{\mu_j} (k_j)$ whose action is defined as follows:
$$
\eqalignno{
\widehat{D}_{\mu_j}
  &(k_j)
   {i\over\slash{p}-m}
  = {i\over \slash{p} -m}
  \gamma_{\mu_j}
   {i\over \slash{p}+\slash{k}_j-m},\cr
\widehat{D}_{\mu_j}
      &(k_j) {i\over p-m} \gamma_{\mu_1}{i\over \slash{p}
+\slash{k}_1-m}\cr
&={i\over \slash{p} -m}
      \gamma_{\mu_j}
     {i\over \slash{p} +\slash{k}_j  - m}\gamma_{\mu_1}
{i\over \slash{p} + \slash{k}_j + \slash{k}_1 -m},\cr
&+ {i\over \slash{p} - m}
     \gamma_{\mu_1}
     {i\over \slash{p}+\slash{k}_1-m}
      \gamma_{\mu_j}
   {i\over \slash{p} + \slash{k}_j+\slash{k}_1-m},\cr
\hbox{etc.} & \mbox{ }\cr
&&(2.9) \cr}
$$
That is,
$\widehat{D}_{\mu_j}(k_j)$
acts on any generalized propagator by
inserting in all possible ways an interaction with a photon of momentum
$k_j$
and vector index
$\mu_j$.
Then one may define
$$
\widehat{Q}_{\mu_j}
(k_j) =
\widehat{D}_{\mu_j}
(k_j) -
\widehat{C}_{\mu_j}
(k_j).\eqno(2.10)
$$
Then the result in momentum space of inserting in all possible ways
(i.e., in all possible orders) into any generalized propagator $P$ of
the kind illustrated in (2.1) a set of $J$ quantum interactions and a
set of $J'$ classical interactions is
$$
\prod_{j'\epsilon J'}
\widehat{C}_{\mu_j'}
(k_{j'})
\prod_{j\epsilon J}
\widehat{Q}_{\mu_j}
(k_j) P.\eqno(2.11)
$$
Consideration of $(2.3)$ and $(2.9)$ shows that the operators $\widehat{C}_i$
and $\widehat{Q}_i$ appearing in $(2.11)$ all commute, provided we reserve
until the
end all integrations over the variables $\lambda_i$, in order for the action of
the operators $\widehat{D}_i$ to be well defined.

One may not wish to combine the results of making insertions in all
orders.
To obtain the result of inserting the classical interaction at just one
place, identified by the subscript $j\epsilon\{1,\cdots ,n\}$,
into a (generalized)
propagator $P_{\mu_1}\cdots _{\mu_n} (p; k_1, \cdots , k_n )$,
abbreviated now by $P_{\mu_j}$,
one begins as in (2.4) with $k_j^{\sigma_j}$ $P_{\sigma_j}$ in place of
the quantity appearing in the bracket.
However, one does not introduce (2.2),
which led to the restriction of the integration to the range $1\geq
\lambda_j \geq 0$.
Then, provided $k_j^2 \neq 0$,
equation (2.4a) gives for the result in momentum space the result
produced by the action of $$
\eqalignno{
\widetilde{C}_{\mu_j}(k_j)&\equiv\cr
&\int^\infty_{0} d\lambda_j O(p_i\to p_i+ \lambda_jk_j
)\left(-{\partial\over \partial p^{\mu_j}}\right)&(2.12)\cr}
$$
upon $k_j^{\sigma_j}P_{\sigma_j}$.

For $k_j^2\neq 0$ this integral converges at the upper endpoint.
The indefinite integral can then be defined so that it vanishes at
$\lambda =\infty$.
We define $\widetilde{C}_{\mu_j}(k_j)$ at $k_j^2=0$ by then using
uniformly only the contribution from the lower endpoint $\lambda =0$, as
was entailed from the start by the initially finite value of $\epsilon$
in (2.4). (Strictly speaking, one should use a Pauli-Villars regulator to
define
the integral in p space---then no special treatment is needed for
$k_j^2=0$)

To obtain a form analogous to (2.12) for the quantum interaction one may
use the identity
$$
\eqalignno{
k_j^{\rho_j} &\int^\infty_0 d\lambda_j \left(-{\partial\over \partial
p^{\rho_j}}\right) P_{\mu_j}(p+\lambda_jk_j)\cr
&= \int^\infty_0 d\lambda_j \left(-{\partial\over
\partial\lambda_j}\right) P_{\mu_j}(p+\lambda_j k_j)\cr
&= P_{\mu_j} (p).&(2.13)\cr}
$$
Then the momentum--space result produced by the insertion of a quantum
coupling in $P_{\mu_1\cdots \mu_n}(p; k_1,\cdots k_\mu) = P_{\mu_j}$
at the vertex identified by $\mu_j$ is generated by the action of
$$
\widetilde{Q}_{\mu_j}(k_j) \equiv (\delta_{\mu_j}^{\sigma_j}k_j^{\rho_j}
- \delta_{\mu_j}^{\rho_j}
k_j^{\sigma_j})\widetilde{C}_{\rho_j}(k_j)\eqno(2.14)
$$
upon $P_{\sigma_j}$ .

An analogous operator can be applied for each quantum interaction.
Thus the generalized momentum--space propagator represented by a line $L$
of $G$
into which $n$ quantum interactions are inserted in a fixed order is
$$
\eqalignno{
&P_{\mu_1\cdots \mu_n}(p; Q, k_1,Q, k_2, \cdots Q, k_n)=\cr
&\prod^n_{j=1} \left[ \int^\infty_0
d\lambda_j(\delta_{\mu_j}^{\sigma_j}k_j^{\rho_j}-
\delta^{\rho_j}_{\mu_j} k_j^{\sigma_j})\left(
-{\partial\over \partial p^{\rho_j}}\right)\right]\cr
&\Big({i\over
\slash{p}+\slash{a}-m}\gamma_{\sigma_1}{i\over\slash{p}+\slash{a} +
\slash{k}_1-m} \gamma_{\sigma_2}{i\over
\slash{p}+\slash{a}+\slash{k}_1+\slash{k}_2-m}\cr
&\cdots \times \gamma_{\sigma_n}{i\over
\slash{p}+\slash{a}+\slash{k}_1+\cdots \slash{k}_n -
m}\Big),&(2.15)\cr}
$$
where
$$
a= \lambda_1 k_1 + \lambda_2 k_2 + \cdots \lambda_n k_n.\eqno(2.16)
$$

If some of the inserted interactions are classical interactions then the
corresponding factors
$(\delta_{\mu_j}^{\sigma_j}k_j^{\rho_j} -
\delta_{\mu_j}^{\rho_j}k_j^{\sigma_j})$ are replaced by
$(\delta_{\mu_j}^{\rho_j}k_j^{\sigma_j})$.

These basic momentum--space formulas provide the starting point for our
examination of the analyticity properties in momentum space,
and the closely related question of infrared convergence.

One point is worth mentioning here.
It concerns the conservation  of  charge condition
$k^\mu J_\mu (k) =0$.
In standard Feynman quantum electrodynamic this condition is not
satisfied by the individual photon--interaction vertex, but is obtained
 only by summing over all the different positions where the photon
interaction can be coupled into a graph.
This feature is the root of many of the difficulties that arise in
quantum electrodynamics.

Equation (2.14) shows that the conservation -- law property holds for the
individual {\it quantum} vertex: there is no need to sum over
different positions.
The classical interaction, on the other hand, has a form that allows one
easily to sum over all possible locations along a generalized
propagator, even before multiplication by $k^\mu$.
This summation converts the classical interaction to a sum of two
interactions, one located at each end of the line associated with the
generalized propagator.
(See, for example, Eq. (7.1) below).
We always perform this summation.Then the classical parts of the interaction
are shifted to the hard--photon interaction points, at which $k^{\mu}J(k)=0$
holds.

\newpage
\noindent{\bf 3. The Quantum Vertex}

Suppose a single quantum interaction is inserted into a line of $G$.
Then the associated generalized propagator is given by (2.11), (2.10), (2.9),
(2.5) and (2.3):
$$
\eqalignno{
P_\mu &(p; \widehat{Q}, k)\cr
&= {i\over \slash{p}-m}
\gamma_\mu {i\over \slash{p}+ \slash{k}-m}\cr
&- \int^1_0 d\lambda {i\over \slash{p}+\lambda\!\!\slash{k}-m}
\gamma_\mu {i\over
\slash{p}+\lambda\!\!\slash{k}-m}.&(3.1)\cr}
$$
The first term in (3.1) is
$$
\eqalignno{
{i\over \slash{p} - m} &\gamma_\mu
   {i\over \slash{p}+\slash{k}-m}\cr
&= - {(\slash{p}+m)\over p^2-m^2} \gamma_\mu {(\slash{p} + \slash{k}+m)\over
(p+ k)^2 - m^2}\cr
&=- (\slash{p} + m) \gamma_\mu (\slash{p}+\slash{k}+m)\cr
&\times \left(
  {1\over p^2 -m^2} \ {1\over 2pk + k^2} - {1\over 2pk + k^2}
\ {1\over (p+k)^2 - m^2}\right)\cr
&=- \bigg[ {-(p^2-m^2) \gamma_\mu + (\sl{p}+m)(2p_\mu +\gamma_\mu\sl{k})\over
(2pk+k^2)(p^2-m^2)}\cr
&- {-((p+k)^2 -m^2)\gamma_\mu + (2p_\mu +2k_\mu -
\sl{k}\gamma_\mu)(\sl{p}+\sl{k}-m)\over (2pk+k^2)(p^2-m^2)}\bigg]\cr
&=- \bigg[{2p_\mu\over (\sl{p}-m)(2pk+k^2)}-{2p_\mu+2k_\mu\over
(\sl{p}+\sl{k}-m)(2pk+k^2)}\cr
&+{1\over (\sl{p}-m)}\times {\gamma_\mu \sl{k}\over
(2pk+k^2)}+{\sl{k}\gamma_\mu\over(2pk+k^2)}\times {1\over
(\sl{p}+\sl{k}-m)}\bigg],&(3.2)\cr}
$$
where $\{\gamma_\mu , \slash{p}\}_+ = 2 p_\mu$ has been used, and $pk$
represents $pk + i0$.

The second term in (3.1) can be computed from standard integral tables.
Then it can be cast into a form similar to (3.2) by first considering it
to be a function of the variable $t=p^2-m^2$,
with $pk$ and $k^2$ regarded as parameters, next separating it into its
meromorphic and nonmeromorphic parts in this variable $t$,
and finally evaluating its meromorphic part as a sum of poles times
residues.
This gives for the meromorphic part
$$
\eqalignno{
\bigg[ \int^1_0 d\lambda
&{1\over \slash{p} + \lambda\!\!\sl{k}-m}
\gamma_\mu
 {1\over \sl{p} + \lambda\!\!\sl{k}-{m}}
 \bigg]_{Mero}\cr
&=\left[
{(\sl{p}+m)\gamma_\mu (\sl{p}+m)\over 2pk~({p}^2-m^2)}
 - {(\sl{p}+\sl{k}+m)\gamma_\mu(\sl{p}+\sl{k}+m)\over
2(p+k)k~((p+k)^2-m^2)}\right]_{Mero}\cr
&= {2p_\mu\over 2pk~(\sl{p}-m)}
-
{2p_\mu+2k_\mu\over 2(p+k)k~(\sl{p}+\sl{k}-m)},&(3.3)\cr}
$$
where a term
not depending on $(p^2-m^2)$ has been dropped from the last line.

The singularities of this function at $pk=0$ and $({p}+k)k=0$ are artifacts
of the separation into meromorphic and non meromorphic parts: their
sum does not have singularities at generic points on these surfaces.
Thus we may replace $pk$ by $pk+i0$ in both the meromorphic and non meromorphic
parts and introduce the identities
$$
{1\over 2pk} = {1\over 2pk+k^2}\left(1+{k^2\over 2pk}\right)\eqno(3.4a)
$$
and
$$
{1\over 2pk+2k^2} =
{1\over 2pk+k^2}
\left(1-{k^2\over 2pk+2k^2}\right).
\eqno(3.4b)
$$
Then the combination of (3.2) and (3.3) gives
$$
\eqalignno{
P_\mu&(p; Q,k)_{Mero}\cr
&= {1\over 2pk+k^2}
\Bigg[{1\over \sl{p}-m}({2p_\mu k^2\over
2pk}-\gamma_\mu\sl{k})\cr
&+ \left( {(2p_\mu+2k_\mu)k^2\over 2pk+2k^2} - \sl{k}\gamma_\mu\right)
{1\over\sl{p}+\sl{k}-m}
\bigg].&(3.5)\cr}
$$
This function is of zeroth order in $\abs{k}$,
whereas the individual contributions
(3.2) and (3.3) are each of order $\abs{k}^{-1}$.

The result (3.3) can be obtained also directly by inspection of the integral
appearing on the left--hand side, written in the form
$$
\int^1_0
d\lambda
{(\sl{p}+\lambda\sl{k}+m)
\gamma_\mu (\sl{p}+\lambda
\sl{k}+m)\over (p^2-m^2 +2pk\lambda + k^2\lambda^2)^2}.
$$
The singularities of this integral lying along the surface $p^2=m^2$ arise
from the endpoint $\lambda =0$ of the domain of integration. Thus the
analytic character of these singularities is controlled by the character
of the integrand in an arbitrarily small neighborhood of this endpoint.
Positive powers of $\lambda$ in the numerator diminish the contributions
from this endpoint, and lead to singularities on $p^2=m^2$ that are, in
form,
not as strong as the singularity coming from the terms that are of zeroth
order in $\lambda$.
Thus to find the strongest singularity we may set the $\lambda$'s
appearing in the numerator to zero.
For similar reasons we can set the $\lambda^2$ terms in the denominator
equal to zero, provided the coefficient $2pk$ of the first power of $\lambda$
is nonzero.
Thus the strongest singularity of the integral arising from the lower endpoint
is
$$
\eqalignno{
\int^\infty_0 d\lambda&{(\sl{p}+m)\gamma_\mu (\sl{p}+m)\over (p^2 -m^2 +
2pk\lambda )^2}\cr
&= {(\sl{p} + m)\gamma_\mu (\sl{p} +m)\over 2pk(p^2 -m^2)}.&(3.6)\cr}
$$

This is just the result obtained from the full calculation.
The other term in (3.3) comes from the other endpoint, $\lambda =1$.
Because the strongest or dominant singularities coming from the two
endpoints are poles any other singularities coming from these endpoints belong
to the nonmeromorphic part.

The full nonmeromorphic part of $P_\mu (p; Q, k)$ is, by direct calculation,
$$
\eqalignno{
P_\mu &(p; Q, k)_{NonMero}\cr
&= \bigg[(\sl{p} +m) \gamma_\mu (\sl{p}+m)
 \left({-2k^2\over -d}\right)\cr
&+ ((\sl{k} \gamma_\mu (\sl{p} + m) +(\sl{p} + m)\gamma_\mu\sl{k})\left(
{2pk\over -d}\right)\cr
&+ \sl{k}\gamma_\mu \sl{k}
\left({-2(p^2-m^2)\over -d}\right) \bigg]\cr
&\times \bigg[{ 1\over\sqrt{-d}} \log \left({1-{\sqrt{-d}\over 2pk+2k^2}\over
1+{\sqrt{-d}\over 2pk+2k^2}}\right)\cr
&- {1\over \sqrt{-d}} \log \left({1- {\sqrt{-d}\over 2pk}\over
1+{\sqrt{-d}\over2pk}}\right)+ {2\over 2pk + 2k^2} -{2\over
2pk}\bigg]&(3.7)\cr}
$$
where $-d=(2pk)^2 -4k^2(p^2-m^2)=(2(p+k)k)^2-4k^2((p+k)^2-m^2)$.
The two non--log terms in the final square bracket cancel the pole singularity
in $t=p^2-m^2$ at $d=0$ that would otherwise arise from the small$-d$ behavior
of the log terms.

\begin{figure}
\caption{
 The singularities of $P_\mu (p; Q, k)$ are confined to the surfaces
$p^2-m^2=0$, $(p+k)^2 - m^2=0$, and the branch of $d=0$ lying between $pk=0$
and $pk =-k^2$.}
\epsfxsize = 6.35in
\epsfysize = 4.53in
\epsffile{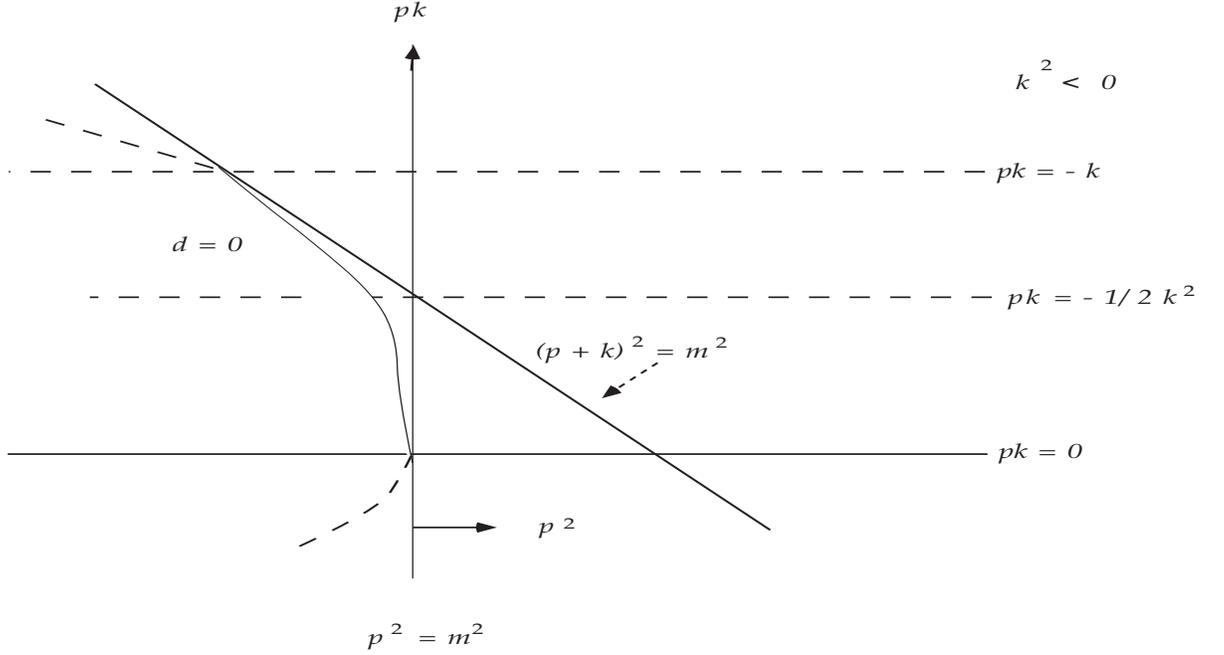}
\label{fig1}
\end{figure}

The singularity surfaces of $P_\mu (p; Q,k)$ are shown in Figure \ref{fig1}.

The singularities of $P_\mu (p;Q,k)$  are confined to the surfaces $p^2-m^2=0,
(p+k)^2 -m^2=0$, and to the portion of the surface $d=0$ that lies between
$pk=0$ and $pk=-k^2$.
Except at points of contact between two of these three surfaces the function
$P_\mu(p; Q,k)$ is analytic on the three surfaces $2pk=0, 2pk+k^2=0$, and
$2pk+2k^2=0$,
and has the form $d^{-3/2}$ on the singular branch of the surface $d=0$.
It has both pole and logarithmic singularities on the surfaces $p^2-m^2=0$
and $(p+k)^2-m^2=0$.
The $i0$ rule associated with $d=0$ matches the $i0$ rules at $p^2=m^2$
and $(p+k)^2=m^2$ at their points of contact.

The meromorphic and nonmeromorphic parts of $P_\mu(p; Q,k)$ each separately
have singularities on the surfaces $2pk=0$, $2pk+k^2=0$ and $2pk+2k^2=0$.

The results of this section may be summarized as follows: the insertion of
a single quantum interaction into a propagator $i(\sl{p}-m)^{-1}$ associated
with $G$ converts it into a sum of three terms.
The first is a propagator $i(\sl{p}-m)^{-1}$ multiplied by a factor that
is zeroth order in $r=\abs{k}$.
The second is a propagator $i(\sl{p}+\sl{k}-m)^{-1}$ multiplied by a factor
that is zeroth order in $r$.
The third is a vertex--type term, which has logarithmic singularities on
the two surfaces $p^2-m^2=0$ and $(p+k)^2-m^2=0$.
This latter term has a typical vertex--correction type of analytic structure
even though it is represented diagrammatically as (the nonmeromorphic part
of) a simple vertex insertion.
\newpage
\noindent{\bf 4. Triangle--Diagram Process}

\begin{figure}
\caption{ Graph representing a soft--photon correction to a hard--photon
triangle--diagram process.
Hard and soft photons are represented by dashed and wiggly lines,
respectively.}
\epsfxsize = 4.52in
\epsfysize = 5.89in
\epsffile{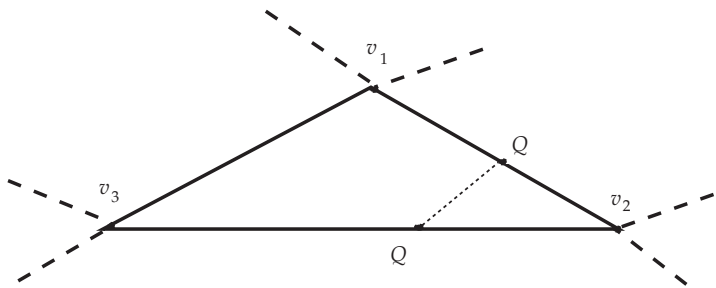}
\label{fig2}
\end{figure}

In the introduction we described a hard--photon process associated with
a triangle graph $G$.
In this section we describe the corrections to it arising from a single
soft photon that interacts with $G$ in the way shown in Figure \ref{fig2}.

Each external vertex $v_i$ of Fig. 1 represents the \ul{two}
vertices upon which the two external hard photons are incident, together
with the charged--particle line that runs between them.
The momenta of the various external photons can be chosen so that the
momentum--energy of this connecting charged--particle line is far from the
mass shell, in the regime of interest.
In this case the associated propagator is an analytic function.
We shall, accordingly, represent the entire contribution associated with
each external vertex $v_i$
by the single symbol $V_i$,
and assume only that the corresponding function
is analytic in the regime of interest.
The analysis will then cover also cases outside of
quantum--electrodynamics.

In Fig. 2 the two solid lines with Q--vertex insertions represent generalized
propagators.
We consider first the contributions that arise from the meromorphic or
pole contributions to these two generalized propagators.

\begin{figure}
\caption{
 Graphs representing the four contributions that arise from inserting
into each of the two generalized propagators represented in Fig. 2 the sum
of the two meromorphic terms given by (3.5).}
\epsfxsize = 6.35in
\epsfysize = 4.53in
\epsffile{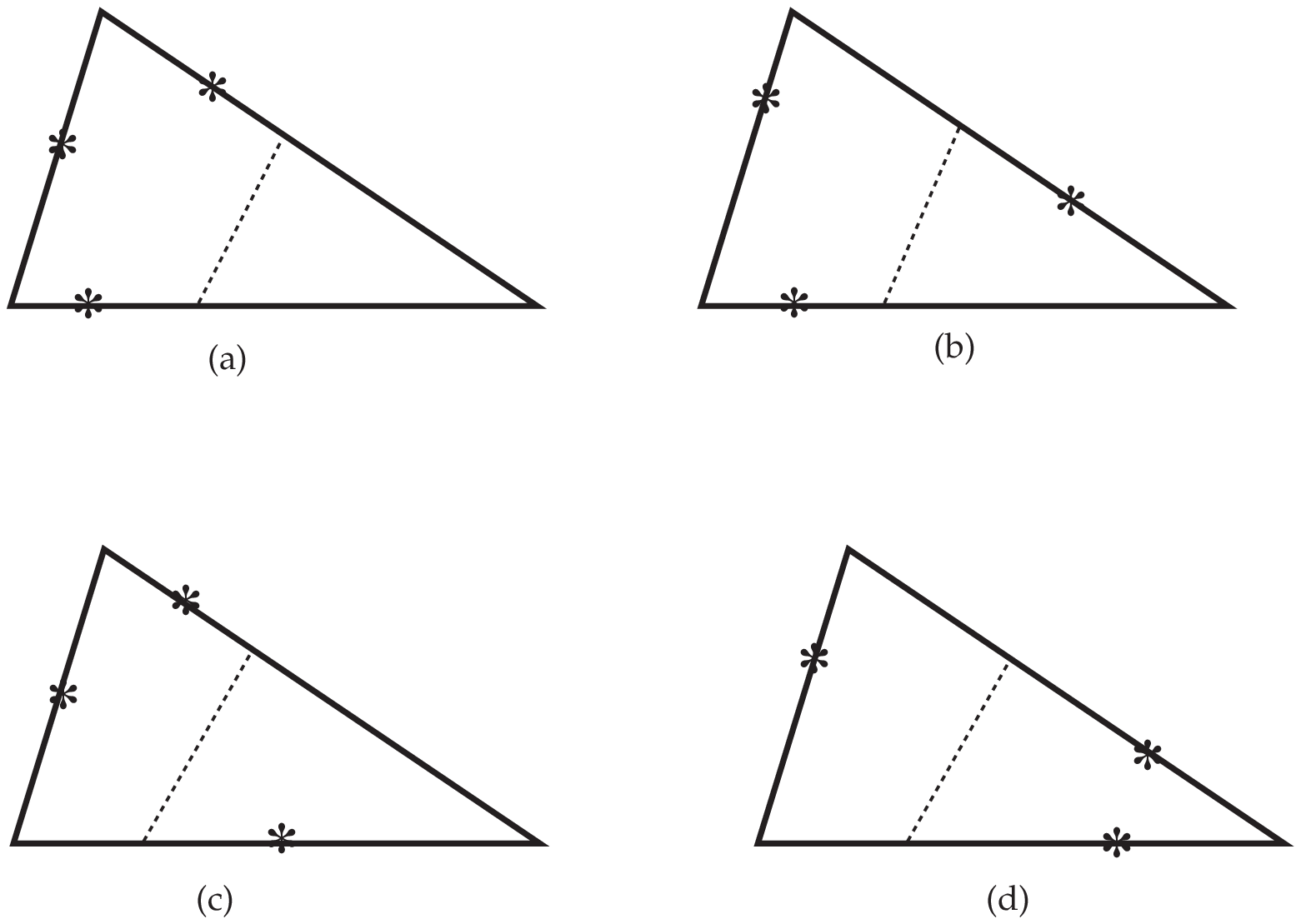}
\label{fig3}
\end{figure}

Each generalized propagator has, according to (3.5), two pole contributions,
one proportional to the propagator $i(\sl{p}-m)^{-1}$, the other
proportional to\newline
 $i(\sl{p}+ \sl{k} -m)^{-1}$.
This gives four terms, one corresponding to each of the four graphs in
Fig. 3.
Each line of Fig. 3 represents a propagator $i(\sl{p}_i-m)^{-1}$ or $i(\sl{p}_i
+\sl{k}-m)^{-1}$ , with $i=1$ or 2 labelling the two relevant lines.
The singularities on the Landau triangle--diagram surface $\varphi =0$
arise from a conjunction of three such singularities, one from each side
of the triangle in Figure \ref{fig3}.

\noindent The graph (a) represents, by virtue of (3.5), the function
$$
\eqalignno{
F_a &= \int {d^4p\over (2\pi)^4}
   \int_{\abs{k}\leq \delta}
   {d^4k\over (2\pi)^4}
i(k^2+i0)^{-1}\cr
&{\rm Tr}\bigg\{{i(\sl{p} +m)\over p^2 -m^2} V_1
{(\sl{p}_1+m)\over p^2_1-m^2}
\bigg( {2p_{1\mu}k^2(2p_1k)^{-1}-\gamma_\mu\sl{k}\over 2p_1 k+k^2}\bigg)\cr
&V_2\left({2p_{2\mu}k^2(2p_2k)^{-1} -\sl{k}\gamma_{\mu 2}\over 2p_2 k+k^2}
\right)
{(\sl{p}_2+m)\over p^2_2 - m^2}V_3\bigg\}&(4.1)\cr}
$$
where $p_1=p+q_1, \ p_2=p-q_3, \ p_ik=p_ik+i0$, and $q_i$ is the
momentum--energy
carried out of vertex $v_i$ by the external hard photons incident upon it.
The vector $p \equiv p_3$ is the momentum--energy flowing along the internal
line that runs from $v_1$ to $v_3$.

To give meaning to the function $(k^2 +i0)^{-1}$ at the point $k=0$ we
introduce polar coordinates, $k=r\Omega$, and write
$$
\int_{\abs{k}\leq\delta} {d^4k\over k^2+i0} f(k) = \int^\delta_0 2rdr\int
d^4\Omega {\delta (\Omega^2_0 + \vec{\Omega}^2-1)\over \Omega^2 + i0}
f(r\Omega)\eqno(4.2)
$$
Then $F_a$ becomes
$$
\eqalignno{
F_a &= \int {d^4p\over (2\pi )^4} \int^\delta_0 2rdr \int
{d^4\Omega\over (2\pi )^4}
{i\delta (\Omega^2_0 +\vec{\Omega}^2-1)\over \Omega^2+i0}\cr
&{\rm Tr}\bigg\{ {i(\sl{p}+m)\over p^2-m^2} V_1 {(\sl{p}_1+m)\over p^2_1-m^2}
\left({2p_{1\mu}\Omega^2 (2p_1\Omega )^{-1} - \gamma_\mu \sl{\Omega}\over
2p_1\Omega + r\Omega^2}\right)\cr
&V_2 \left( {2p_{2\mu}\Omega^2 (2p_2\Omega )^{-1}-\sl{\Omega}\gamma_\mu\over
 2p_2\Omega + r \Omega^2}\right)
{(\sl{p}_2+m)\over p^2_2 -m^2} V_3\bigg\}.&(4.3)\cr}
$$
where $p_i\Omega$ represents $p_i\Omega + i0$.

The integrand of this function behaves near $r=0$
like $rdr$.
Hence the integral is infrared finite.

We are interested in the form of the singularity at interior points of the
positive--$\alpha$ branch of the Landau triangle--diagram surface $\varphi
=0$.
Let $q= (q_1, q_2, q_3)$ be such a point on $\varphi (q)=0$.
The singularity at $q$ is generated by the pinching of the contour of
integration in $p$--space by the three surfaces $p_i^2-m^2=0$.
This pinching occurs at a point in the domain of integration where the three
vectors $(p_1, p_2, p_3)$ lie at a point $(p_1(q), p_2(q), p_3(q))$ that
is determined uniquely by the value $q$ on $\varphi = 0$.
At this point none of these vectors is parallel to any other one.
Consequently, in view of the $i0$ rules described in connection with Fig.
1., it is possible, in a sufficiently small $p$--space neighborhood of $(p_1
(q), p_2(q), p_3(q)$), for sufficiently small $\delta$, to shift the contour
of integration in $\Omega$ space simultaneously into the regions Im $p_1\Omega
>0$ and  Im $p_2\Omega >0$,
and to make thereby the denominator factors $p_i\Omega$ and $p_i\Omega
+r\Omega^2$,for $i\in \{1,2\}$,
all simultaneously nonzero, for all points on the $\Omega$ contour.
In this way the factors in (4.3) that contain these denominator functions
can all be made analytic in all variables in a full neighborhood of the
pinching point.
Consequently, these factors can, for the purpose of examining the character
of the singularity along $\varphi =0$ be incorporated into the analytic
factor $V_2$.

The computation of the form of the singularity on $\varphi =0$ then reduces
to the usual one: the singularity has the form $\log \varphi$, and the
discontinuity is given by the Cutkosky rule, which instructs one to replace
each of the three propagator--poles $i(p^2_i - m^2)$ by $2\pi \delta (p^2_i
- m^2)$.

This gives most of what we need in this special case:
it remains only to be shown that the remaining
singularities on $\varphi =0$ are weaker in form than log$\varphi$.

If one were to try to deal in the same way with the function represented
by Fig. 2, but with the original vertices $\gamma_\mu$ rather than $Q_\mu$,
then (3.2) would be used instead of (3.5) and the integration over $r$ in
the expression replacing (4.3) would become infrared divergent.
The definition of $k^2 +i0$ embodied in (4.2) is insufficient in this case.
A proper treatment$^{10}$ shows that the dominant singularity
on the surface $\varphi =0$ would in this case be $(\log \varphi )^2$.

The graph (b) of Fig. 3 represents the function
$$
\eqalignno{
F_b &= \int {d^4 p\over (2\pi )^4}
  \int^\delta_0 2rdr \int
  {d^4\Omega\over(2\pi )^4} \ \
  {
  i\delta (\Omega^2_0 + \vec{\Omega}^2-1)\over \Omega^2 + i0
   }\cr
&\Tr \Bigg\{ {i(\sl{p}+m)\over p^2-m^2} V_1
\bigg(
  {(2p_{i\mu}+2r\Omega_\mu )
 \Omega^2(2p_1\Omega + 2 r\Omega^2)^{-1} -
\slash{\Omega} \gamma_\mu\over 2p_1 \Omega + r \Omega^2}\bigg)\cr
&\times {(\sl{p}_1 + r\sl{\Omega} + m)\over (p_1+ r\Omega )^2-m^2}
  V_2\cr
&\times \bigg(
  {
    2p_{2\mu}
    \Omega^2(2p_2\Omega)
    -\gamma_\mu \sl{\Omega}\over 2p_2\Omega + r \Omega^2
  }
   \bigg)
 {(\sl{p}_2 +m)\over p^2_2 - m}
  V_3 \bigg\}
&(4.4)\cr}
$$
where $p_i\Omega$ represents $p_i\Omega + i0$.
This integral also is free of infrared divergences.
It is shown in ref. 15
  that its singularity on $\varphi =0$ has the form
$\varphi^2\log\varphi$.
The same result is obtained for graphs (c) and (d) of Fig. 3.

The remaining contributions to the process represented in Fig. 2 involve
the nonmeromorphic parts of at least one of the two generalized propagators.
These nonmeromeorphic parts are given by (3.7).
This expression gives logarithmic singularities on $p^2_i - m^2=0$ and $(p_i
+ r\Omega )^2 - m^2=0$, for $i=1$ and 2.
It gives singularities also on $p_i\Omega = 0$ and $p_i\Omega + r \Omega^2
=0$, and a $d_i^{-3/2}$ singularity on the portion of the surface $d_i=0$
that lies between $p_i\Omega =0$ and $p_i\Omega + r \Omega^2 =0$.

For $p$ in a small neighborhood of the fixed pinching point one can again,
for sufficiently small $\delta$, distort the $\Omega$ contour simultaneously
into the upper--half planes of both $p_1\Omega$ and $p_2\Omega$, and thereby
avoid simultaneously the zeros of $p_i\Omega$, $p_i\Omega = r\Omega^2$,
and also those of
$$
d_i (2r)^{-2} = (p_i\Omega )^2 - (p_i^2 - m^2)\Omega^2.$$
Thus for every point on the $\Omega $ contour the nonmeromorphic part of
the propagator associated with line $i$ takes, near the pinching point,
the form
$$
A_i {1\over r} \log {(p_i + r\Omega )^2 - m^2\over p^2_i - m^2},\eqno(4.5)
$$
where $A_i$ is analytic in all variables.

If we combine the two factors (4.5), one from each end of the photon line,
then the two displayed powers of $r^{-1}$ join with $rdr$ to give $dr/r$.
Consequently, if each of the two logarithmic factors in (4.5) were treated
separately then an infrared divergence would ensue.
However, the entire (4.5), taken as a unit, is of zeroth order in $r$, and
it gives no such divergence.
It is therefore necessary in the treatment of the nonmeromorphic part to
keep together  those contributions coming from various logarithmic
singularities, such as the two logarithmic singularities of (4.5),
 that are naturally tied together by a cut.
By contrast,
in the meromorphic part it was possible to treat separately the contributions
from the two different pole singularities associated with each of the two
sides $i=1$ and $i=2$ of the triangle: for the meromorphic part each of
the four terms indicated in Fig. 3 is separately infrared convergent.

The product of the two factors (4.5) gives an integrand factor of the form
$$
I={dr\over r} (\log {(p_1+r\Omega )^2-m^2\over p^2_1 - m^2}) (\log {(p_2+
r\Omega )^2 - m^2\over p^2_2 - m^2}).\eqno(4.6)
$$
The dominant singularity on $\varphi =0$ generated by this combination is
shown in ref. 15 to be of the form $\varphi^2(\log \varphi )^2$.
If one combines the nonmeromorphic part from one end of the soft--photon
line with the meromorphic part from the other end then the resulting dominant
singularity on $\varphi =0$ has the form $\varphi \log \varphi$.
Replacement of {\it{one}} of the two $Q$--type interactions in Fig. 2 by
a $C$--type interaction does not materially change things.
The results are described in ref. 15.

We now turn to the generalization of these results to processes involving
arbitrary numbers of soft photons, each having a $Q$--type interaction
on at least one end.
\newpage
 \noindent{\bf 5. Residues of Poles in Generalized Propagators}

Consider a generalized propagator that has only quantum--interaction
insertions.
Its general form is, according to (2.15),
$$
\eqalignno{
\prod^{n}_{j=1}&\left[\left(
\delta_{\mu_j}^{\sigma_j} k^{\rho_j}_j -
\delta_{\mu_j}^{\rho_j} k^{\sigma_j}_j \right)
\int^\infty_0 d\lambda_j \left(-{\partial\over\partial p^{\rho_j}}
\right)\right]\cr
&( {i\over \sl{p} +\sl{a}-m}
\gamma_{\sigma_1}
{i\over \sl{p} +\sl{a}+\sl{k}_1-m}
\gamma_{\sigma_2}
{i\over\sl{p}+\sl{a}+\sl{k}_1+\sl{k}_2-m}\cr
&\cdots \times \gamma_{\sigma_{n}}
{i\over \sl{p}+ \sl{a} + \sl{k}_1 \cdots
+ \sl{k}_n -m} \bigg)&(5.1)\cr}
$$
where
$$
a=\lambda_1 k_1 + \cdots + \lambda_n k_n .\eqno(5.2)
$$

The singularities of (5.1) that arise from the multiple end--point $\lambda_1
= \lambda_2 = \cdots \lambda_n =0$ lie on the surfaces
$$
p^2_i = m^2,\eqno(5.3)
$$
where now (in contrast to earlier sections)
$$
p_i = p + k_1 + k_2 +\cdots +  k_i.\eqno(5.4)
$$
At a point lying on only one of these surfaces the strongest of these
singularities is a pole.
As the first step in generalizing the results of the preceding section to
the general case we compute the residues of these poles.

The Feynman function appearing in (5.1) can be decomposed into a sum of
poles times residues.
At the point $a=0$ this gives
$$
\eqalignno{
&{i(\sl{p}+m)\gamma_{\mu_1} i(\sl{p}+\sl{k}_1+m)
\gamma_{\mu_2}\cdots
\gamma_{\mu_n} i(\sl{p}+\cdots +
\sl{k}_n+m)\over
(p^2-m^2) ((p+k_1)^2-m^2)((p+\cdots + k_n)^2-m^2)}\cr
& \ \ \ \ \ = \sum^n_{i=0} {N_{1i}\over D_{1i}}
{i(\sl{p}_i+m)\over p^2_i-m^2}
{N_{2i}\over D_{2i}},&(5.5)\cr}
$$
where for each $i$ the numerator occurring on the right--hand side of this
equation is identical to the numerator occurring on the left--hand side.
The denominator factors are
$$
D_{1i} = \prod_{j<i} (2p_i k_{ij}+(k_{ij})^2 + i0),\eqno(5.6a)
$$
and
$$
D_{2i} = \prod_{j>i}(2p_i k_{ij} + (k_{ij})^2 + i0),\eqno(5.6b)
$$
where
$$
\ k_{ij} = \sigma_{ij} [(k_1 + \cdots + k_j)-(k_1 + \cdots + k_i)].\eqno(5.7)
$$

     The sign $\sigma_{ij}=\pm$ in (5.7) is specified in the following way: in
order to
make the pole-residue formula well defined each quantity $p_sk_i$ is replaced
by $p_sk_i+i\epsilon_i$ with $\epsilon_i>>\epsilon_{1+1}>0$, for the ordering
(6.1). Thus each $\epsilon_i$ is taken to be much larger than the next one,
so it that it dominates over any sum of smaller ones. This makes each
difference
of denominators that occurs in the pole-residue decomposition well defined,
with a well-defined nonvanishing imaginary part. Then the sign $\sigma_{ij}$
in (5.7), is fixed so as to make the imaginary part of the $(i,j)$ factor in
(5.6) positive. Then the limit where all $\epsilon_i \rightarrow 0$ is
concordant with (5.6).

Since the singularities in question arise from the multiple endpoint
$\lambda_1=\cdots \lambda_n =0$ it is sufficient for the  determination
of the analytic character of the singularity to consider an arbitrarily
small neighborhood of this endpoint.
We shall consider, for reasons that will
be explained later, only points in a closed domain in the variables $k_j$
upon which the parameters $p_ik_j$ and $2p_i k_{ij} + (k_{ij})^2$ are all
nonzero.
Then the factors $D_{1i}^{-1}$ and $D^{-1}_{2i}$ are analytic functions
of the variables $\lambda_j$ in a sufficiently small neighborhood of the
point $\lambda_1 = \cdots = \lambda_n=0$.
Hence a power series expansion in these variables can be introduced.

The dominant singularity coming from the multiple end point $\lambda_1 =
\cdots = \lambda_n =0$ is obtained by setting to zero all the $\lambda_j$
coming from either the numerators $N_{1i}$ and $N_{2i}$ or the power series
expansion of the factors $D_{1i}^{-1}$ and $D_{2i}^{-1}$.
Then the only remaining $\lambda_j$'s are those in the pole factor $((p_i
+ a)^2 -m^2)^{-1}$ itself.

Consider, then, the term in (5.1) coming from the {\it i\/}th term in (5.5).
And consider the action of the first operator, $j=1$, in (5.1).
This integral is essentially the one that occurred in section 3.
Comparison with (2.3), (3.6), and (3.3) shows that the dominant singularity
on  $p^2_i - m^2 =0$ is the function obtained by simply making the replacement
$$
\int^\infty_0 d\lambda_j \left( - {\partial\over \partial p^{\rho_j}}\right)
\left( O(p\to p+\lambda_j k_j)\right) \to p_{i\rho_j}(p_ik_j)^{-1}.\eqno(5.8)
$$
Each value of $j$ can be treated in this way.
Thus the dominant singularity of the generalized propagator (5.1) on $p_i^2
- m^2=0$ is
$$
\eqalignno{
   \prod^n_{j=1}
&\left[ \left( \delta_{\mu_j}^{\sigma_j} k_j^{\rho_j} - \delta_{\mu_j}^{\rho_j}
k_j^{\sigma_j}\right)
p_{i\rho_j} (p_i k_j)^{-1}\right]\cr
 &\times {N_{1i} i(\sl{p}_i+m)N_{2i}\over
D_{1i}(p^2_i-m^2)D_{2i}}.
&(5.9)\cr}
$$

The numerator in (5.9) has, in general, a factor
$$
\eqalignno{
&\ \ \ \ \ i(\sl{p}_i- \sl{k}_i+m)\gamma_{\sigma_i}i(\sl{p}_i+m)
\gamma_{\sigma_{i+1}}i(\sl{p}_i+\sl{k}_{i+1}+m)\cr
   &=i(\sl{p}_i-\sl{k}_i+m)\gamma_{\sigma_i} i((\sl{p}_i+m)i(2p_{i\sigma_{i+1}}
+ \gamma_{\sigma_{i+1}}\sl{k}_{i+1})\cr
&\ \ \ \ \  +i(\sl{p}_i - \sl{k}_i +m) \gamma_{\sigma_i}
\gamma_{\sigma_{i+1}}(p^2_i-m^2)\cr
&=i(2p_{i\sigma_i}-\sl{k}_i\gamma_{\sigma_i})i(\sl{p}+m)i(2p_{i\sigma_{i+1}}
+ \gamma_{\sigma_{i+1}}\sl{k}_{i+1})\cr
&\ \ \ \ \
+i(p^2_i-m^2)\gamma_{\sigma_i}(2p_{i\sigma_{i+1}}+\gamma_{\sigma_{i+1}}
\sl{k}_{i+1})\cr
&\ \ \ \ \
+i(\sl{p}_i-\sl{k}_i+m)\gamma_{\sigma_i}\gamma_{\sigma_{i+1}}(p^2_i-m^2)
&(5.10)\cr}
$$

The last two terms in the last line of this equation have factors
$p^2_i-m^2$.
Consequently, they do not contribute to the residue of the pole at
$p_i^2-m^2=0$. The  terms in (5.10) with a factor $2p_{i\sigma_{i+1}}$,
taken in conjunction with the factor in (5.9) coming from $j=i+1$, give
a dependence $2p_{i\rho_j} 2p_{i\sigma_j}$.
This dependence upon the indices $\rho_j$ and $\sigma_j$ is symmetric under
interchange of these two indices.
But the other factor in (5.9) is antisymmetric.
Thus this contribution drops out.
The contribution proportional to $p_{i\sigma_i}$ drops out for similar reasons.

Omitting these terms that do not contribute to the residue of the  pole
at $p^2_i - m^2$ one obtains in place of (5.10)
the factor
$$
(-i\sl{k}_i\gamma_{\sigma_i})i(\sl{p}_i+m)(i\gamma_{\sigma_{i+1}}\sl{k}_{i+1})\eqno(5.11)
$$
which is first--order in both $\sl{k}_i$ and $\sl{k}_{i+1}$.

The above argument dealt with the case in which $i\neq 0$ and $i\neq n$:
i.e., the propagator $i$ is neither first nor last.
If $i=0$ then there is no factor $k_i=k_0$ in (5.11): in fact no such $k_j$
is defined.
If $i=n$ then there is no factor $k_{i+1}=k_{n+1}$ in (5.11):
in fact no such $k_j$ is defined in the present context.
Thus one or the other of the two $k-$ dependent factors drops out if propagator
$i$ is the first or last one in the sequence.

This result (5.11) is the generalization to the case $n>1$ of the result for
$n=1$ given in (3.5).
To obtain the latter one must combine (5.11) with (5.9).
The effect of (5.11) is to provide, in conjunction with these pole
singularities,
a ``convergence factor''
for the factors lying on either side of each pole factor in the pole--residue
decomposition (5.5).
That these ``convergence factors''
actually lead to infrared convergence is shown in the following sections.

\newpage
\noindent{\bf 6. Infrared Finiteness of Scattering Amplitudes.}

Let $G$ be a hard--photon graph.
Let $g$ be a graph obtained from it by the insertion of soft photons.
In this section we suppose at that each soft photon is connected on both ends
into $G$ by a $Q$--type interaction.

Each charged--particle line segment $L$ of $G$ is converted into a line
$L'$ of $g$ by the insertion of $n \geq 0$ soft--photon vertices.
The line $L'$ of $g$ represents a generalized propagator.
Let the symbols $L_i$, with $i\in \{0, \ldots n\}$,
represent the various line segments of $L'$.

In this section we shall be concerned only with the contributions coming
from the pole parts of the propagator described in section 5.
In this case each generalized propagator is expressed by (5.9) as a sum
of pole terms, each with a factorized residue enjoying property (5.11).

One class of graphs is of special interest.
Suppose for each charged line $L'$ of $g$ there is a  segment $L_i$
such that the cutting of each of these segments $L_i$,
together perhaps with the cutting of some hard--photon lines, separates
the graph $g$ into a set of disjoint subgraphs each of which contains
precisely one vertex of the original graph $G$.
In this case the soft--photon part of the computation decomposes into several
independent parts:
all dependence on the momentum $k_j$ of the soft photon $j$ is confined
to the functional representation of the subgraph in which the line
representing this photon is contained.

The purpose of this section is first to prove infrared convergence for the
special case of {\it separable} graphs defined by two conditions.
The first condition is that the graph $g$ separate into subgraphs
in the way just described.
We then consider for each line $L'$ of $g$ a single term in the
corresponding generalized propagator (5.9).
The second condition is that  in this term of (5.9) the factor $i(\sl{p}_i+m)
(p^2_i-m^2)^{-1}$ correspond to the line segment of $L_i$ that is cut to
produce
the separation into subgraphs.
Then each subgraph will contain, for each charged--particle
line that either enters it or leaves it, a
half--line $h$
that contains either the set of vertices $j\geq i$, or, alternatively, the
set of vertices $j<i$, of that charged--particle line.

It is also assumed that the graph $G$ is simple: at most one line segment
(i.e., edge) connects any pair of vertices of $G$.

The contributions associated with graphs of this kind are expected to
give the dominant singularities of the full function on the Landau surface
associated with $D$.
If the functions associated with all the various subgraphs are well defined
when the momenta associated with all lines of $D$ are placed on--mass--shell
then the discontinuity of the full function across this Landau surface will
be a product of these well defined functions.
By virtue of the spacetime fall--off
properties established in paper I these latter
functions can then be identified with contributions to the scattering functions
for processes involving charged external particles.
The purpose of this section is to prove the infrared finiteness of these
contributions to the  scattering functions.

Each subgraph can be considered separately.
Thus it is convenient to introduce a new labelling of the set of, say, $n$
soft photons that couple into the subgraph under consideration.
To do this the domain of integration
$0 \leq \abs{k_j} \leq \delta , \ j \in\{ 1, \ldots , n\}$, is first decomposed
into $n$!
domains according to the relative sizes of the Euclidean magnitudes
$\abs{k_j}$.
Then in each of these separate domains the vectors $k_i$ are labelled so that
$\abs{k_1} \geq \abs{k_2}\geq \ldots \geq \abs{k_n}\geq 0$.
A generalized polar coordinate system is then introduced:
$$
  \eqalignno{
  k_1 &= r_1\Omega_1\cr
  k_2 &= r_1r_2\Omega_2\cr
\vdots \ \ \ &\cr
k_n &= r_1r_2\cdots r_n \Omega_n.
&(6.1)\cr}
$$
Here $\abs{r_1}\leq \delta$, and $\abs{r_j}\leq 1$ for $j=2, \cdots n$,
and $\Omega \tilde{\Omega}\equiv (\Omega_{j0})^2+(\vec{\Omega}_j)^2=1$.

The factors in $D_i (a=0)$, as defined in (5.6), are $2p_i k_{ij} +
(k_{ij})^2$. However, the $k_{ij}$ are no longer given by (5.7).
With our new labelling the formula (5.7) becomes
$$
k_{ij} = \sum_{j'\in J(i,j)} \pm \  k_{j'},\eqno(6.2)
$$
where the signs $\pm$ are the same as the signs in (5.7): only the labelling of
the vectors is changed.

Let $j(i, j)$ be the smallest number in the set of numbers $J(i, j)$.
Then singling out this term in $k_{ij}$ one
may write
$$
2p_ik_{ij}+ (k_{ij})^2 = r_1r_2 \cdots r_{j(i,j)}(2p_i\Omega_{j(i,j)}+R)
\eqno(6.3)
$$
where $R$ is
bounded.

The zeros of the factors $(2p_i\Omega_{j(i,j)}+R)$
play an important role in the integration over $\Omega $ space.
However, our objective in this section is to prove the convergence of the
integrations over the radial variables $r_j$,
under the condition that the $\Omega$ contours can be distorted so as to
keep all of these $\Omega$--dependent factors finite, and hence analytic.
The validity of this distortion condition  is discussed in Section 8, and
proved in ref. 14.

To prove infrared convergence under this condition it is sufficient to
show, for each value of $j$, that if the differential $dr_j$ is considered
to be of degree one in $r_j$ then the full integrand, including the
differential $dr_j$,
is of degree at least two in $r_j$.
This will ensure that the integration over $r_j$ is convergent near $r_j
=0$.

The power counting in the variables $r_{j'}$ is conveniently performed in
the following way:
the factor $\abs{k_j} d\abs{k_j}$ arising from $d^4k_j/k^2_j+i0$ gives,
according
to (6.1), a factor that has, in each variable $r_{j'}$,
the degree of $(r_1\cdots r_j)^2$.
This factor may be separated into two factors $(r_1\cdots r_j)$,
one for each end of the photon line.
Then each individual generalized propagator can be considered separately:
for each coupling of a photon $j$ carrying momentum $k_j=r_1\cdots r_j\Omega_j$
into a half--line $h$ we assign to $h$ one of the two factors $(r_1\cdots
r_j)$ mentioned above.
Thus each half--line $h$ will have one such numerator factor for each of the
photon lines that is incident upon it, and this numerator factor can be
associated
with the vertex upon which the photon line is incident.
On the other hand, (6.3) entails that there is a dominator factor $r_1\cdots
r_{j(i,j)}$ associated with the $j$th interval of $h$.
Finally, if the photon incident upon the endpoint of $h$ that stands next to
the interval
that was cut is labelled by $e$ then there is an extra numerator factor
$r_1\cdots r_e$:
it comes from the factor $\sl{k}_{i+1}$ (or $\sl{k}_i$) in (5.11).

We shall now show that these various numerator and denominator factors combine
to produce for each $j$, and for each half--line upon which the soft photon
$j$, is incident, a
net degree  in $r_j$ of at least one, and for every other half--line a
net degree of at least zero.

Consider any fixed $j$.
To count powers of $r_j$ we first classify each soft photon $j'$ as
``nondominant'' or ``dominant'' according to whether $j'\geq j$ or $j' <
j$.
Any line segment of $h$ along which flows the momentum $k_{j'}$ of a dominant
photon $j'$ will, according to (6.3), not contribute a denominator factor
$r_j$.

Thus the denominator factors that do contribute  a power of  $r_j$ can be
displayed
graphically by first considering the line $h$ that starts at the initial
vertex $j=e$, which stands, say, just to the right of the cut line--segment
$L_i$, and that runs to the right.
Soft photons are emitted from the succession of vertices on $h$, and some
of these photons can be reabsorbed further to the right  on $h$.
In such cases the part of $h$ that lies to the right of the vertex where
a dominant photon is emitted but to the left of the point where it is
reabsorbed may be contracted to a point: according to (6.3) none of these
contracted line segments of $h$ carry a denominator factor of $r_j$.
If a dominant soft photon is emitted but is never reabsorbed on $h$ then
the entire part of the line $h$ lying to the right of its point of emission
can be contracted to this point.

If the line obtained by making these two changes in $h$ is called $h'$ then,
by virtue of (6.3), there is exactly one denominator factor $r_j$ for
each line segment of $h'$.

Self--energy and vertex corrections are to be treated in the usual way by
adding counterterms.
Thus self--energy--graph insertions and vertex--correction graphs should
be omitted: the residual corrections do not affect the power counting.
This means that every vertex on $h'$, excluding the last one on the right
end, will be either:
\begin{enumerate}
\item An original vertex from which a single nondominant photon is either
emitted or absorbed; or
\item A vertex formed by a contraction.
Any vertex of the latter type must have at least two nondominant soft photons
connected to it, due to the exclusion of self--energy and vertex corrections.
\end{enumerate}

The first kind of vertex will contribute one power of $r_j$ to the numerator,
whereas the second kind of vertex will contribute at least two powers of
$r_j$.

Every line segment of $h'$ has a vertex standing immediately to its left.
Thus each denominator power of $r_j$ will be cancelled by a numerator power
associated with this vertex.
This cancellation ensures that each half--line will be of degree at least
zero in $r_j$.

If the soft--photon $e$ incident upon the left--hand end of $h'$ is nondominant
then one extra power of $r_j$ will be supplied by the factor $\sl{k}_e$
coming from (5.11).
If the soft photon $e$ is dominant then there are two cases:
either the left--most vertex of $h'$ is the only vertex on $h'$, in which
case there are no denominator factors of $r_j$, but at least one numerator
factor for each $k_j$ vertex incident on $h$; or the left most vertex of
$h'$ differs from the rightmost one, and is formed by contraction, in which
case at least two nondominant lines  must be connected to it.
These two lines deliver two powers of $r_j$ to the numerator and hence the
extra power needed to produce degree one in $r_j$.

This result for the individual half lines means that for the full subgraph
the degree in $r_j$ is at least one for every $j$.
Hence the function  is infrared convergent.

The argument given above covers specifically only the special class of
separable
graphs $g$. However, the argument applies essentially unchanged to the
general case. The restriction to separable graphs fixed the directions that
the photon loops flowed along the half-line $h$ under consideration: each
photon loop $i$ incident upon $h$ flowed {\it away} from the pole
line-segment $s$ that lies on one end of $h$. This entails that for any line
segment $j$ lying in $h$ the associated denominator function $f_j$ contains
a term $2p_sk_i$ if and only if the following condition is satisfied: exactly
one end of the photon loop $i$ that carries momentum $k_i$ is incident upon
the half-line $h$ in the interval lying between the (open) segment $j$ and the
(open) segment $s$ that lies on the end of $h$.

This key property of $f_j$ follows in general, however, directly from the
formula
$$
\eqalignno{
  f_j & = \sigma_{js} (\Sigma^2_j - \Sigma^2_s) \cr
      & = \sigma_{js} (\Sigma_j + \Sigma_s)(\Sigma_j - \Sigma_s) \cr
      & = \sigma_{js} (2p_s(\Sigma_j - \Sigma_s)+\Sigma^2_j - \Sigma^2_s),
&(6.4) \cr}
$$
where $\Sigma_j=p_s+ K_j$ and $\Sigma_s= p_s + K_s$. The difference $K_j-K_s$
consists, apart from signs, of the sum of the $k_i$ associated with the
photon loops $i$ that are incident upon $h$ precisely {\it once} in the
interval between the segments $j$ and $s$. This entails the key property
that was obtained in the separable case from the separability condition, which
is consequently not needed: the arguments in this section pertaining to the
powers of the $r_i$ cover also the non-separable case.

\newpage
\noindent{\bf 7. Inclusion of the Classical Interactions}

The power--counting arguments of the preceeding section dealt with processes
containing only $Q$--type interactions.
In that analysis the order in which these $Q$--type interactions were inserted
on the line $L$ of $G$ was held fixed: each such ordering was considered
separately.

In this section the effects of adding $C$--type interaction are considered.
Each $C$--type interactions introduces a coupling $k^\sigma\gamma_\sigma
=\sl{k}$.
Consequently, the Ward identities, illustrated in (2.7),
can be used to simplify the calculation, but only if the contributions
from all orders of its insertion are treated together.
This we shall do.
Thus for $C$--type interactions it is the  operator $\ha$
defined in (2.5) that is to be used rather than the
operator $\wt$ defined in (2.12).

Consider, then, the generalized propagator obtained by inserting on some line
$L$
of $G$ a set of $n$ interactions of $Q$--type, placed in some definite
order, and a set of $N$ $C$--type interactions, inserted in all orders.
The meromorphic part of the function obtained after the action of the
$n$ operators $\wwt_j$ is given by (5.9).
The action upon this of the $N$ operators $\ha_j$ of (2.5) is obtained by
arguments similar to those that gave (5.9), but differing by  the fact
that (2.5) acts upon the propagator present {\it before\/} the action of
$\ha_j$,
and the fact that now both limits of integration contribute, thus giving for
each $\ha_j$ two terms on the right--hand side rather than one.
Thus the action of $N$ such $\ha_j$'s gives $2^N$ terms:
$$
\eqalignno{
\Bigg[
   \prod^{n+N}_{j=n+1}   &\ha_{\mu_j}(k_j)
   P_{\mu_1\cdots \mu_n}
   (p; Q, k_1, Q, k_2, \cdots Q, k_n)
   {\Bigg]}_{Mero}\cr
&= \sum^{2^N}_{\Theta=1}
  S{gn}(\Theta )\sum^n_{i=0}
  \prod^{n+N}_{j=n+1}
  \left(
  {ip^\Theta_{i \mu_j}\over p^\Theta_i k_j}
  \right)\cr
&\times
  \left\{ \prod^n_{j=1}
  \left[
  \left(
  \delta_{\mu j}^{\sigma_j}k_j^{\rho_j}
  - \delta^{\rho_j}_{\mu_j}k_j^{\sigma_j}
  \right)
  \left(
  {p^\Theta_{i\rho_j}\over p^\Theta_ik_j}
  \right)
  \right]
  \right\}
  \cr
&\times
  {N^\Theta_{1i}\over D^\Theta_{1i}}
  {i(\sl{p}^\Theta_i+m)\over (p^\Theta_i)^2-m^2}
  {N^\Theta_{2i}\over D^\Theta_{2i}},
&(7.1)
\cr}
$$
where
$$
\eqalignno{
\Theta &= (\Theta_{n+1}, \cdots , \
\Theta_{n+N}),\cr
\Theta_j &= +1 \  {\hbox{or}} \ 0,\cr
S{gn}(\Theta )&= (-1)^{\Theta_{n+1}}(-1)^{\Theta_{n+2}}\cdots
(-1)^{\Theta_{n+N}}\cr
p_i^\Theta &= p_i + \Theta_{n+1}k_{n+1} +\cdots + \Theta_{n+N}k_{n+N},\cr
p_i &= p+k_1 + \cdots + k_i,
&(7.2)
\cr}
$$
and the superscript $\Theta$ on the $N$'s and $D$'s means that
the argument $p_i$
appearing in (5.5) and (5.6) is replaced by $p_i^\Theta$.
Note that even though the action of $\ha_j$ and $\wwt_j$
involve integrations over $\lambda$ and differentiations, the meromorphic
parts of the resulting generalized propagators are expressed by (7.1) in
relatively simple closed form.
These meromorphic parts turn out to give the dominant contributions in the
mesoscopic regime, as we shall see.

The essential simplification obtained by summing over  all orders of the
$C$--type
insertions is that after this summation each $C$--type interaction gives
just two terms.
The first term is just the function before the action
of $\ha_j$ multiplied by $ip_{i\mu_j} (p_i k_j)^{-1}$; the second is minus
the same thing with $p_i$ replaced by $p_i +k_j$.
Thus, apart from this simple factor, and, for one term, the overall shift
in $p_i$, the function is just the same as it was before the action of
$\ha_j$.
Consequently, the power--counting argument of section 6 goes through
essentially unchanged:
there is for each classical photon $j$ one extra denominator factor $(p_ik_j)$
coming from the factor $ip_{iu_j}(p_i k_j)^{-1}$ just described, but the
powers of the various $r_i$ in this denominator factor are exactly cancelled
by the numerator factor $(r_1\cdots r_j)$ that we have associated with the
vertex $\ha_j$.
Because of this exact cancellation the C-type couplings do not
contribute to the power counting. Hence when C-type couplings are allowed
the arguments of section 6 lead to the result that the
meromorphic part of the function $F$ associated with the quantum photons is of
degree at least  one in each of the
variables $r_j$. Hence it is infrared convergent.

              \newpage
\noindent{\bf 8. Distortion of the $\Omega$ Contours }

The proof of infrared finiteness given in sections 6 and 7 depends upon the
assumption that the $\Omega $ contours can be shifted away from all denominator
zeros in the residue factors of any term in the pole-residue
decomposition of the Feynman function corresponding to the simple triangle
graph, modified by the insertion of an arbitrary number of soft-photon lines,
each of which has a quantum coupling on at least one end and a quantum or
classical coupling on the other. The proof that such a distortion of the
contour is possible requires two
generalizations of the available results about the locations of singularities
occurring in the terms of the perturbative expansion in field theory.

In the first place, we
must deal not only with the Feynman functions themselves, but also with
the functions obtained by decomposing, according to the pole-residue theorem,
the generalized propagators associated with the three sides of the triangle.
For the usual Feynman functions themselves there is available the useful
geometric formulation, in terms of Landau diagrams, of necessary conditions
for a singularity.  In ref. 14 we have developed a generalization of the
Landau-diagram condition that covers the more general kinds of functions that
arise in our work.

The second needed generalization pertains to the masslessness
of photons. If the standard Landau-diagram momentum-space conditions
are generalized to include massless particles then the effect of contributions
from points where $k_i=0$, for some $i$, is to produce a severe weakening of
the necessary conditions. But in ref. 14 the needed strong results are obtained
in the variables $(r_i, \Omega_i)$ introduced in section 6 to prove infrared
finiteness.
\newpage

\noindent{\bf 9. Contributions of the Meromorphic Terms
 to the Singularity on the Triangle-Diagram Surface $\varphi =0$.}
\vskip 9pt

In this section we describe the contributions to the singularity on the
triangle-diagram singularity surface $\varphi =0$ arising
from the meromorphic parts of the three generalized propagators.

The arguments of sections 6, 7, and 8 show that in the typical pole-residue
term (5.9) we can distort the contours in the $\Omega_j$
variables so as to keep the  residue factors analytic, even in the
limit when some or all of the $r_j$'s become zero.
In that argument we considered separately an individual half-line, but the
argument is `local': it carries over to the full set of six half-lines, with
all the $|k_i|$ ordered.
Thus for each fixed value of the set of variables $(r_i,...,r_{n}; \
\Omega_1,..., \Omega_n)$ the integration over the remaining variable of
integration $p$ gives essentially a triangle-graph function: it gives a
function with the same log $\varphi$-type singularity that arises from the
simple Feynman triangle-graph function itself, with, however, the location of
this
singularity in the space of the external variables $(q_i, q_2, q_3)$ shifted
by an amount $(K_1, K_2, K_3)$, where the three vectors $K_s$ are related to
the photon momenta flowing along the three star lines of the original
graph.
Specifically, if we re-draw the photon loops so that they pass through {\it
no} star line of the original graph (or equivalently through no star line of
the Landau diagram), but pass, instead, {\it out} of the graph at a vertex
$v_1, v_2$ or $v_3$, if necessary, and then define the net momentum flowing
out of vertex $v_s$ to be
$$
q_s =q_s (k) +K_s,\eqno(9.1)
$$
where $K_s$ is the net momentum flowing out of vertex $v_s$ along the newly
directed photon loops, then, for fixed $k$, the function in $(q_1, q_2, q_3)
$ space will have a normal log $\varphi $ triangle-diagram singularity
along the surface $\varphi (q_1(k), q_2(k), q_3(k)) =0$.
For example, the original singular point at the point $\hat{q}$ in
$(q_1, q_2, q_3) $ space will be shifted to the point
$(q_1, q_2, q_3) = (\hat{q_1},\hat{q_2},\hat{q_3})+(K_1, K_2, K_3)$.
This shift in the external variables $q$'s shifts the momentum flowing
along the three star lines to the values they would have if the photon
moments $k_i$ were all zero: it shifts the kinematics back to the one where
no photons are present.

It is intuitively clear that the smearing of the location of this log
$\varphi$ singularity caused by the integration of the variables $k_i$ will
generally produce a weakening of the log singularity at $\varphi (q) =0$.
For, in general, only the endpoint $r_1 =0$ of the $r_1$ integration will
contribute to the singularity at $\varphi (q) =0$, and there is no divergence
at $r_1=0$, by power counting, and hence no contribution from this set of
measure zero in the domain of integration.
The only exception arises from the set of separable graphs.
For in these graphs the
 $K_s$ are all zero, and hence the integrations produce no smearing, and
thus no weakening, of the log $\varphi$ singularity.

To convert this intuitive argument to quantitative form we begin by
separating the set of photon lines into two subsets that enter differently into
the calculations. Let a {\it bridge} line in a graph $g$ that corresponds to
a term in the pole-residue decomposition $(7.1)$ be a photon line $j$ that
`bridges' over a star line: any closed loop in $g$ that contains the
photon line segment $j$, and is completed by charged-particle segments that
lie on the triangle $G$, passes along at least one star line. Let $i$ be the
smallest $j$ such that photon line $j$ is a bridge line. (Here we are using the
ordering of the full set $(1,2, \ldots, n)$ of photon labels that was specified
in $(6.1)$, not the ordering used in $(7.1)$). Thus each
$k_j= \rho_j \Omega_j=r_1...r_j\Omega_j$ that appears in a star-line
denominator, and hence in $(9.1)$, contains a factor $\rho_i=r_1...r_i$.
Let the set of variables $(k_1,...,k_{i-1})$ be denoted by $k_a$, and let the
set of variables $(k_i,...,k_n)$ be denoted by $k_b$. And let $r_a$ and $r_b$,
and $\Omega_a$ and $\Omega_b$ be defined analogously.  Then the function
represented by $g$ can be written in the form
$$
F(q)=\prod^n_{j=i}\int_{\Omega_j\widetilde{\Omega}_j=1}d\Omega_j
\int^1_0 r_j^{e_j} dr_j\ G(q,\Omega_b,r_b) \eqno(9.2)
$$
where
$$
\eqalignno{
G(q,\Omega_b,r_b)=&\int d^4p\ \prod^{i-1}_{j=1}
\int_{\Omega_j\widetilde{\Omega}_j=1} d\Omega_j
\int^1_0 r_j^{e_j} dr_j \cr
&\prod^3_{s=1}{1\over{p_s(q,\Omega_b,r_b)^2-m^2=+i0}}\ R(q,\Omega_b,r_b,
\Omega_a,r_a).&(9.3)\cr}
$$
Here R is the product of the three residue factors.

The integrations in (9.2) weaken the logarithmic singularities: it is shown in
ref. 15 that the singularity on the surface $\varphi (q) =0$ is contained in
a finite sum of terms of the form $A_m\varphi$ (log $\varphi)^m$, where $m$ is
a
positive integer that is no greater than the number of photons in the
graph, and $A_m$ is analytic.

\newpage

\noindent{\bf 10. Operator Formalism.}
\vskip 9pt

We have dealt so far mainly with the meromorphic contributions.
In order to treat the nonmeromorphic remainder it is convenient to decompose
the operator $\widehat{C}_i$ into its ``meromorphic  and ``nonmeromorphic''
parts, $\widehat{C}^M_i$ and $\widehat{C}^N_i$.

The operator $\widehat{C}_i$ is defined in (2.5):
$$ i \widehat{C}_iF(\widetilde{p}) = \int^1_0 d\lambda_i {\partial\over
\partial
p^{\mu_i}} F(p)\eqno(10.1)
$$
where
$$
p=\widetilde{p} + \lambda_i k_i.\eqno(10.2)
$$
Suppose
$$
F(p) = A(p)B(p), \eqno(10.3)
$$
where $A(p)$ is analytic and $B(p)$ is $(p^2-m^2)^{-1}$.
An integration by parts gives
$$
\eqalignno{
i\widehat{C}_i AB &= \int^1_0 d\lambda_i \bigg[(\partial_{\mu_i}A)B +
A(\partial_{\mu_i}B)\bigg]\cr
&= \int^1_0 d\lambda_i \Bigg[ (\partial_{\mu_i} A)- (\partial A/\partial
\lambda_i) \int^{\lambda_i}\partial_{\mu_i}\cr
&\ \ +A\left(\delta(\lambda_i-1)-\delta(\lambda_i)\right)\
 \int^{\lambda_i}\partial_{\mu_i}\Bigg]B,
&(10.4)\cr}
$$
where the difference of delta functions, $(\delta(\lambda_i-1)-
\delta(\lambda_i))$ indicates that one is to take the difference
of the integrand at the two end points.

The indefinite integral, computed by the methods used to compute
$(3.3)$, $(3.6)$, and $(3.7)$, is
$$
\eqalignno{
\int^{\lambda_i} & \partial_{\mu_i} B \equiv \int d\lambda_i
{\partial \over\partial p^{\mu_i}}B\cr
&={2p_{\mu_i}\over 2pk_i}B -{4(p_{\mu_i}k^2_i-k_{i\mu_i}pk_i)\over d}
\Bigg[\int^{\lambda_i}B + {1\over pk_i}\Bigg].
&(10.5)\cr}
$$

Because the factor in front of the square bracket in  $(10.5)$ is
independent of  $\lambda_i$ one can use a second integration by
parts (in reverse) to obtain
\newpage
$$
\eqalignno{
i\widehat{C}_i AB &= \cr
\int^1_0 d\lambda_i &\Bigg[\left( \partial_{\mu_i} A\right)
-\left({\partial\over \partial \lambda_i}A\right) {2p_{\mu_i}\over 2pk_i} \cr
&+ A\left(\delta(\lambda_i-1) -\delta(\lambda_i)\right)
{2p_{\mu_i}\over 2pk_i} \cr
&-{4\left(p_{\mu_i}k^2_i- k_{i\mu_i} pk_i \right)\over d} A \cr
&+ {4(p_{\mu_i}k^2_i-k_{i\mu_i}pk_i) \over d} {k^2_i (p^2- m^2)A
\over (pk_i)^2}  \Bigg]B, &(10.6)\cr}
$$
where the final term comes from the $1/pk_i$ term in the square bracket in
$(10.5)$ and has no singularity at $(p^2- m^2)\equiv B^{-1}=0$ for
$pk_i \neq 0$.

Since all of the $\lambda_i$ dependence in A is in $p = \tilde{p} +
\lambda_ik_i$
we may write
$$
\partial A/\partial\lambda_i = (\partial_{\mu_i}A) k^{\mu_i}.\eqno(10.7)
$$
Hence the first two terms on the right side of (10.6) cancel, and one is left
with
$$
\wh{C}_i = \wh{C}^M_i + \wh{C}^N_i + \wh{C}^R_i,\eqno(10.8)
$$
where
$$
i\wh{C}^M_i AB = \int^1_0 d\lambda_i {2p_{\mu i}\over
2pk_i}(\delta(\lambda_i-1)
-\delta(\lambda_i)) AB \eqno(10.8a)
$$
$$
i\wh{C}^N_i AB = -{4(p_{\mu_i} k^2_i - k_{\mu_i} pk_i)\over d} \int^1_0
d\lambda_i
AB\eqno(10.8b)
$$
$$
\eqalignno{
i\wh{C}^R_i AB &= {4(p_{\mu_i}k^2_i - k_{\mu i} p k_i)\over d} \int^1_0
d\lambda_i \Bigg[\left({\partial\over\partial\lambda_i}A\right)\cr
& -
A(\delta (\lambda_i-1)-\delta (\lambda_i))\bigg] {1\over pk_i} \cr
&= {4(p_{\mu_i}k^2_i-k_{i\mu_i}pk_i)\over d} \int^1_0 d\lambda_i {k^2_i
(p^2- m^2)\over(pk_i)^2}AB.&(10.8c)\cr}
$$
Notice that the contribution $\wh{C}^R_i$ cancels the pole at $d=0$ of the
contribution $\wh{C}^N_i$.

To efficiently manipulate these operators their commutation relations are
needed.
Recall from section 2 that the operators $\widehat{C}_i$ commute among
themselves,
as do the $\widehat{D}_i$:
$$
[\widehat{C}_i, \widehat{C}_j]=0\eqno(10.9a)
$$
and
$$
[\widehat{D}_i, \widehat{D}_j]=0.\eqno(10.9b)
$$
The operators $\widehat{C}_i$ and $\widehat{D}_j$, properly interpreted, also
commute:
$$
[\widehat{C}_i, \widehat{D}_j] =0.\eqno(10.9c)
$$

To verify (10.9c) note first that $\widehat{D}_j$ acts on generalized
propagators
(See (2.9)), and, by linearity, on linear superpositions of such propagators.
However, Eq. (2.3) shows that the action on such an operand of the operator
$(-\partial /\partial p_{\mu_i})$ in $\widehat{C}_i$ is the same as a
$\widehat{D}_i$ with $k_i=0$. Moreover, the replacement $p\to p+\lambda_ik_i$
commutes with $\widehat{D}_j$.
Thus (10.9c) is confirmed, provided we stipulate that the
integrations over the variables $\lambda_i$ shall be reserved until the end,
{\it after} the actions of all operators $\widehat{D}_i$ and differentiations.
In fact, we see from $(10.8)$ that the various partial operators $\wh{C}^M_i$,
$\wh{C}^N_j$, and $\wh{C}^R_k$ all commute: if we reserve the $\lambda$
integrations until the end then each of the operations is implemented by
multiplying the integrand by a corresponding factor, and
those operations commute.
\newpage
\noindent{\bf 11. Nonmeromorphic Contributions}

The $D$-coupling part of a $Q$-type coupling is meromorphic. Thus
each of the $\widehat{C}$- and $\widehat{Q}$-type couplings can
be expressed as by means of $(10.8)$ as sum of of its meromorphic,
nonmeromorphic, and residual parts. Then the full function can be expanded
as a sum of terms in which each coupling is either $\widehat{C}$-type or
$\widehat{Q}$-type, and is either meromorphic, nonmeromorphic, or residual.
If any factor is residual then the term has no singularity at $(p^2-m^2)=0$,
and  is not pertinent to the question of the singularity structure on
$\varphi=0$. Thus these residual terms can be ignored.

We have considered previously the terms in which every coupling is meromorphic.
Here we examine the remainder. Thus terms not having least one nonmeromorphic
coupling $\widehat{C}^N_i$ or $\widehat{Q}^N_i$ are not pertinent: they can
also be ignored.

All couplings of the form $\widehat{Q}^M_i$ can be shifted to the
right of all others, and this product of factors $\widehat{Q}^M_i$ can then be
re-expressed in terms of the couplings $\widetilde{Q}^M_i$. That is, the terms
corresponding to the different orderings of the insertions of the
{\it meromorphic} couplings $Q^M_i$ into the charged-particle lines can be
recovered by using $(2.9)$, $(2.15)$, and $(5.8)$. The various couplings
$\widehat{C}^N_i$ are then represented, apart from the factor standing outside
the integral in $(10.8b)$, simply by an integration from zero to one
on the associated variable $\lambda_i$.

In this paper we are
interested in contributions such that every photon has a $Q$-type coupling on
at least one end. In sections 6 and 7 the variables $\rho_i$'s corresponding
to photons $i$ having a $\widetilde{Q}^M_i$-type coupling on (at least) one end
were expressed in
terms of the variables $r_j$, and it was shown that the contributions from all
of the  $\widetilde{Q}^M_i$-type couplings lead to an $r_j$ dependence that is
of
order at least one in each $r_j$. The $\widehat{C}_i^M$-type couplings do not
upset
this result. Thus the general form of the expression that represents any
term in the pole-residue expansion of the product of meromorphic couplings
$\widetilde{Q}^M_i$ and $\widehat{C}^M_i$  is
$$
\prod_j \int_{\Omega_j \widetilde{\Omega}_j=1} d\Omega_j \prod_i
\int^1_0 r^{e_i}_i dr_i\ AB,  \eqno(11.1)
$$
where the $e_i$ are nonnegative integers, and $A$ and $B$ have the forms
specified in section 10, provided the $\Omega$ contours are distorted in the
way described in section 8 and ref. 14. (For convenience, the scale has been
defined so
that the upper limit $\delta $ of the integration over $r_1$ is unity.)

For these meromorphic couplings the integrations over the variables $\lambda_i$
have been eliminated by the  factors $\delta (\lambda_i - 1)$ and $\delta
(\lambda_i)$. But for any coupling $\widehat{C}^N_i$ there will be, in
addition to the integration from zero to one on the variable $r_i$, also an
integration from zero to one on the variable $\lambda_i$. It comes from
$(10.8b)$.

These integrals are computed in ref. 15, and it is shown that the
nonmeromorphic contributions lead to the singularities on the triangle
diagram singularity surface $\varphi=0$ that are no stronger than
$\varphi(\log \varphi)^{n+1}$, where $n$
is the number of photons in the graph.
Even if the log factors from the graphs of different order in $e^2$ should
combine to give a factor like $\varphi^{-(1/137)}$, this factor, when combined
with the form $\varphi(\log \varphi)^{n+1}$, would not produce
a singularity as strong as the $\log\varphi$ singularity that arises from the
separable graphs.

\newpage
\noindent{\bf 12. Comparison to Other Recent Works}
\vskip 9pt

Block and Nordsieck$^{12}$ recognized already in 1937 that a large part of the
very soft photon contribution to a scattering cross-section was correctly
predicted by classical electromagnetic theory. They noted that the process
therefore involves arbitrarily large numbers of photons, and that this
renders perturbation theory inapplicable. They obtained finite results
for the cross section for the scattering of a charged particle by a
potential $V$ by taking the absolute-value squared of the matrix element of $V$
between initial and final states in which each charged particle is
``clothed'' with a cloud of bremsstrahlung soft photons. The two key ideas
of Block and Nordsieck are, first, to focus on a physical quantity, such as
the observed cross section, with a summation over unobserved very soft
photons, and, second, to separate out from the perturbative treatment the
correspondence--principle part of the scattering function, which is also the
dominant contribution at very low energies.

These ideas have been developed and refined in an enormous number of articles
that have appeared during the more than half-century following the paper of
Block and Nordsieck. Particularly notable are the works of J. Schwinger$^{17}$,
Yennie, Frautschi, and Suura$^{18}$, and K.T. Mahanthappa$^{19}$.
Schwinger's work was the first modern treatment of the infrared divergence
problem, and he conjectured exponentiation. Yennie, Frautschi, and Suura,
formulated the problem in terms of Feynman's diagramatic method, and analyzed
particular contributions in detail. They gave a long argument suggesting that
their method should work in all orders, but their argument was admittedly
nonrigorous, and did not lend itself to easy rigorization. The main
difficulties had to do with the failure of their arguments at points where the
basic scattering function was singular. These points are precisely the focus of
the present work, and our way of separating out the dominant parts leads to
remainder terms that are compactly representable, and hence amenable to
rigorous treatment. Mahanthappa considered, as do we, closed time loops,
and split the photons into hard and soft photons, and constructed an electron
Green's function in closed form for the soft-photon part to do perturbation
theory in terms of the hard part.

General ideas from these earlier works are incorporated into the present work.
But our logical point of departure is the article of Chung$^{20}$ and of
Kibble$^4$. Chung was the first to treat the scattering
amplitudes directly, instead of transition probabilities, and to introduce,
for this purpose, the coherent states of the electromagnetic field.
Kibble first exhibited the apparent break-down of the
pole-factorization property in QED. The present work shows that this effect is
spurious: the non-pole form does not arise, at least in the case that we have
examined in detail, if one separates off for nonperturbative treatment not the
approximate representation of the correspondence-principle part used by Chung
and Kibble, but rather an accurate expression that is valid also in
case the scattering process is macroscopic, and that therefore involves no
replacement of factors $\exp ikx$ by anything else.

The works mentioned above are not directly comparable to present one
because they do not address the question at issue here, which is the
large-distance behaviour of quantum electrodynamics, and in particular the
dominance at large distances of a part that conforms to the correspondence
principle and enjoys the pole-factorization
property. The validity of these principles in quantum elecrodynamics
is essential to the logical structure of quantum theory: the relationship
between
theory and experiment would become ill-defined if these principles were to
fail. These principles are important also at the practical level.
The domain of physics lying between the atomic and classical regimes
is becoming increasingly important in technology. We therefore need to
formulate the computational procedures of quantum electrodynamics
in a way that allows reliable predictions to be made in this domain.
Moreover, the related ``problem of measurement'' is attracting increasing
attention among theorists and experimentalists. The subject of this work is
precisely the subtle mathematical properties of this quantum-classical
interface in the physical theory that actually controls it. Finally, the
problem of the
effects of massless particles in gauge theories is an issue of mounting
theoretical importance. Theorists need to have an adequate treatment of
this mathematically delicate problem in our premier physical theory, quantum
electrodynamics, which serves as a model for all others.

Kulish and Faddeev$^{21}$ have obtained a finite form of quantum
electrodynamics by modifing the dynamics of the asymptotic states.
For our purposes it is not sufficient merely to
make the theory finite. We are interested in the  nature of singularities,
and the related question of the  rates of fall-off for large
spacetime separations. To obtain a sufficiently well-controlled computational
procedure, in which no terms with spurious rates of fall off are introduced by
an unphysical separation of the problem into parts, it was important, in our
definition of the classical
part, to place the sources of the classical radiation field, and of the
classical ``velocity'' fields, at their correct locations. The needed
information about the locations of the scattering sites is not naturally
contained in the asymptotic states: the scattering events can involve both
``in'' and ``out'' particles together, and perhaps also internal particles
as well. We bring in the correct locations of the scattering sites by
rearranging the terms of the coordinated-space perturbative expansion of the
full scattering operator itself, rather than by redefining the initial and
final states of the S-matrix.

d'Emilio and Mintchev$^{22}$ have initiated an approach that is
connected to the one pursued here.
They have considered charged-field operators that are nonlocal in that each
one has an extra phase factor that is generated by an infinite line
integral along a ray that starts at the field point $x$.
Their formula applied to the case of a product of three current operators
located at the three vertices $(x_1, x_2, x_3$) of our closed triangular
loop {\it could} be made to yield precisely the phase that appears in Eq.
(1.7) of ref. 11.
However, that would involve making the direction of the ray associated with
each field operator $\psi (x)$ depend upon the argument of the {\it other}
field operator in the coordinate-space Green's function
$\langle T\psi (x) \bar{\psi} (x')\rangle_0$ in which it appears.

d'Emilio and Mintchev do not follow that tack.
Instead, they keep the direction of the ray associated with each field
operator $\psi (x)$ fixed, {\it then} go to momentum space, and then find,
for some simple cases (charged-particle propagation and vertex correction),
that reasonable results are obtained only if the directions of
the rays associated with the charged-particle operators that create or
destroy the electrons are set equal to the momenta of the particles that are
created or destroyed.

Of course, charged-particle propagators generally occur under integral signs,
whereas the directions of the rays are treated as constants.
If these ``constant'' directions are allowed to depend upon the momentum $p$
then the inverse Fourier transform would, of course, not yield the original
coordinate-space Feynman function.

An ``intuitive'' reason was given why the one particular choice of the
directions of the rays gives reasonable answers. It relies on the idea of
``the classical currents responsible for the emission of soft photons''.
But classical-current arguments ought to be formulated in coordinate space.

Such a formulation (i.e., a coordinate-space formulation) would suggest
letting the direction of the ray that occurs in the d'Emilio-Mintchev formula
be the direction of the line between the two arguments $x$ and $x'$ of the
coordinate-space charged-particle propagator.
Then, due to a partial cancellation, the two infinite line integrals would
collapse to a single finite line integral running between the two points $x$
and $x'$. Then, in the case of our triangular closed loop, the phase factors
associated with the lines on the three sides of the triangle would combine to
give just the phase factors appearing in (1.7) of reference 11.

This coordinate-space procedure, which would seem to be the physically
reasonable way to  proceed, would bring the d'Emilio-Mintchev formulation to
the
first stage of the work pursued here and in reference 11.

The problem of formulating quantum electrodynamics in an axiomatic
field-theoretic framework has been examined by Fr\"{o}hlich, Morchio, and
Strocchi$^{8}$ and by D. Buchholz$^9$, with special attention  to the
non-local aspects arising from Gauss' law.  Their main conclusion, as it
relates to the present work, is that the energy-momentum spectrum
of the full system can be separated into two parts, the first being the
photonic asymptotic free-field part, the second being a remainder that: (1), is
tied to charged particles; (2), is nonlocal relative to the photonic part; and
(3), can have a discrete part corresponding to the electron/positron mass. This
separation is concordant with the structure of the QED
Hamiltonian, which has a photonic free-field part and an electron/positron
part that incorporates the interaction term $eA^{\mu} J_{\mu}$, but
no added term corresponding to the non-free part of the electromagnetic field.
It is also in line with the separation of the classical electromagnetic field,
as derived from the Li\'{e}nard-Wiechert potentials, into a ``velocity'' part
that is attached (along the light cone) to the moving source particle, and an
``acceleration'' part that is radiated away. It is the ``velocity''
part, which is tied to the source particle, and which falls off only as
$r^{-1}$,
that is the origin of the ``nonlocal'' infraparticle structure that
introduces peculiar features into quantum electrodynamics, as compared to
simple local field theories.

In the present approach, the quantum analog of
this entire classical structure is incorporated into the formula for the
scattering operator by the unitary factor $U(L)$. It was shown in ref. 11,
Appendix C, that the non-free ``velocity''
part of the electromagnetic field generated by
$U(L)$ contributes in the correct way to the mass of the electrons and
positrons. It gives also the ``Coulomb'' or ``velocity'' part of the
interaction between different charged particles, which is the part of the
electromagnetic field that gives the main part of Gauss' law asymptotically.
Thus our formulas supply in a computationally clean way these ``velocity
field'' contributions that seem so strange when viewed from other points of
view.

\newpage

\noindent{\bf References}

\begin{enumerate}

\item J. Bros {\it in} Mathematical Problems in Theoretical Physics:
Proc. of the Int. Conf. in Math. Phys. Held in Lausanne Switzerland
Aug 20-25 1979, ed. K. Osterwalder, Lecture Notes in Physics 116,
Springer-Verlag (1980);
H. Epstein, V. Glaser, and D. Iagolnitzer, Commun. Math. Phys. {\bf80},
99 (1981).
\item D. Iagolnitzer, {\it Scattering in Quantum Field Theory: The Axiomatic
and Constructive Approaches}, Princeton University Press, Princeton NJ,
in the series: Princeton Series in Physics. (1993);
J. Bros, Physica {\bf 124A}, 145 (1984)
\item D. Iagolnitzer and H.P. Stapp, Commun. Math. Phys. {\bf 57}, 1 (1977);
D. Iagolnitzer, Commun. Math. Phys. {\bf 77}, 251 (1980)
\item T. Kibble, J. Math. Phys. {\bf 9}, 315 (1968);
Phys. Rev. {\bf 173}, 1527 (1968); {\bf 174}, 1883 (1968); {\bf 175}, 1624
(1968).
\item D. Zwanziger, Phys. Rev. {\bf D7}, 1082 (1973).
\item J.K. Storrow, Nuovo Cimento {\bf 54}, 15 (1968).
\item D. Zwanziger, Phys. Rev. {\bf D11}, 3504 (1975); N. Papanicolaou, Ann.
Phys.(N.Y.) {\bf 89}, 425 (1975)
\item J. Fr\"{o}hlich, G. Morchio, and F. Strocchi, Ann.Phys.(N.Y) {\bf 119},
241 (1979); Nucl. Phys. {\bf B211}, 471 (1983); G. Morchio and F. Strocchi,
{\it in} Fundamental Problems in Gauge Field Theory, eds. G. Velo and
A.S. Wightman, (NATO ASI Series) Series B:Physics {\bf 141}, 301 (1985).
\item D. Buchholz, Commun. Math. Phys. {\bf 85}, 49 (1982); Phys. Lett. B
{\bf 174}, 331 (1986); {\it in} Fundamental Problems in Gauge Field Theory,
eds.
G. Velo and A.S. Wightman, (NATO ASI Series) Series B: Physics {\bf 141}, 381
(1985);
\item T. Kawai and H.P. Stapp, {\it in} 1993 Colloque International en
l'honneur de Bernard
Malgrange (Juin, 1993/ at Grenoble)  Annales de l'Institut Fourier {\bf 43.5},
1301 (1993)
\item H.P. Stapp, Phys. Rev. {\bf 28D}, 1386 (1983).
\item F. Block and A. Nordsieck, Phys. Rev. {\bf 52}, 54 (1937).
\item G. Grammer and D.R. Yennie, Phys. Rev. {\bf D8}, 4332 (1973).
\item T. Kawai and H.P. Stapp, {\it Quantum Electrodynamics at Large Distances
II: Nature of the Dominant Singularities.}
Lawrence Berkeley Laboratory Report LBL 35972. Submitted to Phys. Rev.
\item T. Kawai and H.P. Stapp, {\it Quantum Electrodynamics at Large Distances
III: Verification of Pole Factorization the Correspondence Principle},
Lawrence Berkeley Laboratory Report LBL 35973. Submitted to Phys. Rev.
\item T. Kawai and H.P. Stapp, {\it Quantum Electrodynamics at Large
Distances}, Lawrence Berkeley Laboratory Report LBL-25819 (1993).
\item J. Schwinger Phys. Rev. {\bf 76}, 790 (1949).
\item D. Yennie, S. Frautschi, and H. Suura, Ann. Phys. (N.Y.) {\bf 13},
379 (1961).
\item K.T. Mahanthappa. Phys. Rev. {\bf 126}, 329 (1962); K.T Mahanthappa and
P.M. Bakshi, J. Math. Phys. {\bf 4}, 1 and 12 (1963).
\item V. Chung, Phys. Rev. {\bf 140}, B1110 (1965)
\item P.P. Kulish and L.D. Fadde'ev, Theor. Math. Phys. {\bf 4}, 745 (1971).
\item E. d'Emilio and M. Mintchev, Fortschr. Phys. {\bf 32}, 473 (1984).

\end{enumerate}

\end{document}